\documentclass[12pt]{article}

\usepackage{dsfont}
\usepackage{fullpage}
\usepackage{subfig}
\usepackage{amsmath}
\usepackage{amssymb}
\usepackage{wasysym}
\usepackage{bbm}
\usepackage{color}
\usepackage{ulem}

\newcommand{\Comment}[1]{{}}
\definecolor{MyDarkBlue}{rgb}{0.15,0.15,0.45}
\usepackage[linktocpage=true]{hyperref}
\hypersetup{
colorlinks=true,
citecolor=MyDarkBlue,
linkcolor=MyDarkBlue,
urlcolor=MyDarkBlue,
pdfauthor={},
pdftitle={},
pdfsubject={hep-th}
}

\DeclareFontFamily{OT1}{rsfs10}{}
\DeclareFontShape{OT1}{rsfs10}{m}{n}{ <-> rsfs10 }{}
\DeclareMathAlphabet{\mathscript}{OT1}{rsfs10}{m}{n}

\def\gsim{ \lower .75ex \hbox{$\sim$} \llap{\raise .27ex \hbox{$>$}} }
\def\lsim{ \lower .75ex \hbox{$\sim$} \llap{\raise .27ex \hbox{$<$}} }
\def\be{\begin{equation}}
\def\ee{\end{equation}}
\def\bea{\begin{eqnarray}}
\def\eea{\end{eqnarray}}

\newcommand{\ns}{\normalsize}

\def\({\left(}
\def\){\right)}

\newcommand{\half}{\frac{1}{2}}

\newcommand{\Mpl}{M_{\rm Pl}}
\newcommand{\minf}{m_\mathrm{inf}}

\usepackage{latexsym,amsmath,amssymb,epsfig}

\usepackage{graphicx}
\usepackage{epstopdf}
\usepackage[body={17.5cm, 21cm},right=2cm]{geometry}
\usepackage{amssymb}
\usepackage{amsmath}
\usepackage{psfrag}
\usepackage{epsfig}
\usepackage{cancel}
 \allowdisplaybreaks[4]

\usepackage[all]{xy}

\begin{document}

\begin{titlepage}

%\vspace{-3cm}

\title{
  \hfill{\ns }  \\
  %[1em]
   {\LARGE Symmetron Cosmology}
\\
%[1em] 
}
\author{
   Kurt Hinterbichler, Justin Khoury, Aaron Levy and Andrew Matas
     \\
     %[0.5em]
{\ns Center for Particle Cosmology, Department of Physics \& Astronomy} \\[-0.25cm]
{\ns  University of Pennsylvania, Philadelphia, PA 19104}\\
[0.3cm]}

\date{}

\maketitle

\begin{abstract}
The symmetron is a scalar field associated with the dark sector whose coupling to matter
depends on the ambient matter density. The symmetron is decoupled and screened in
regions of high density, thereby satisfying local constraints from tests of gravity, but
couples with gravitational strength in regions of low density, such as the cosmos.
In this paper we derive the cosmological expansion history in the presence of a symmetron
field, tracking the evolution through the inflationary, radiation- and
matter-dominated epochs, using a combination of analytical approximations and numerical integration.
For a broad range of initial conditions at the onset of inflation, the scalar field reaches its symmetry-breaking vacuum by the present epoch, as assumed
in the local analysis of spherically-symmetric solutions and tests of gravity. For the simplest form of the potential, the energy scale is too
small for the symmetron to act as dark energy, hence we must add a cosmological constant to drive late-time cosmic acceleration. 
We briefly discuss a class of generalized, non-renormalizable potentials that can have a greater impact on the late-time cosmology,
though cosmic acceleration requires a delicate tuning of parameters in this case. 
\end{abstract}

\end{titlepage}

\section{Introduction}

The standard $\Lambda$CDM cosmology explains cosmic acceleration with a cosmological constant, a logically sound but seemingly highly tuned explanation.  Therefore there has been much interest in exploring the possibility that the cosmic acceleration could be caused by a previously unobserved dynamical component of the universe~\cite{Ratra:1987rm}$-$\cite{Das:2005yj}. At the same time, nearly massless gravitationally coupled scalars are generically predicted to exist by many theories of high energy physics.  No experimental sign of such a fundamental scalar has yet been seen, in spite of tests designed to detect solar system effects and fifth forces that would generally be expected if such scalars were present~\cite{Fischbach:1999bc,Will:2005va}.  

If light scalars exist, they must utilize a screening mechanism~\cite{Khoury:2010xi} to hide themselves from local experiments.
Screening mechanisms rely on non-linearities whose behavior depends on the ambient matter density. In regions of high density,
such as the local environment, the scalars develop non-linearities that effectively decouple them from matter. In regions of low density, such as the cosmos,
the scalars couple to matter with gravitational strength and mediate a long-range force, thereby affecting the nature of gravity
and the growth of structure on large scales~\cite{Jain:2010ka}.

Aside from cosmology, these theories find independent motivation in the vast experimental effort aimed at testing the fundamental nature of gravity at long wavelengths~\cite{Will:2005va}. Viable screening theories make novel predictions for local gravitational experiments. The subtle nature of these signals have forced experimentalists to rethink the implications of their data~\cite{eotvos,Upadhye:2006vi} and have inspired the design of novel experimental~\cite{Gies:2007su}$-$\cite{admx} and observational tests~\cite{claire}$-$\cite{Jain:2011ji}. The theories of interest offer a rich spectrum of testable predictions, from laboratory to extra-galactic scales. 

Only a handful of successful and robust screening mechanisms have been proposed to date~\cite{Khoury:2010xi}. The first is the Vainshtein mechanism~\cite{vainshtein,ags,ddgv}, which works when the scalars have derivative self-couplings which become important in the vicinity of massive sources such as the Earth. The strong coupling boosts the kinetic terms, so that after canonical normalization the coupling of fluctuations to matter is weakened.  This mechanism is central to the phenomenological viability of brane-world modifications of gravity~\cite{Dvali:2000hr}$-$\cite{Agarwal:2011mg},
massive gravity~\cite{ddgv},~\cite{Babichev:2009us}$-$\cite{Koyama:2011xz} (see \cite{Hinterbichler:2011tt} for a review), degravitation models~\cite{Dvali:2002pe}$-$\cite{Patil:2008sp} and galileon theories~\cite{galileon}$-$\cite{Goon:2011qf}. See~\cite{Afshordi:2008rd}$-$\cite{Brax:2011sv} for some phenomenological implications.

A second screening mechanism is the chameleon mechanism~\cite{Khoury:2003aq}$-$\cite{Brax:2010gi}, which works when the scalars are non-minimally coupled to matter in such a way that their effective mass depends on the local matter density. Deep in space, where the local mass density is low, the scalars are light and display their effects, but near the Earth, where experiments are performed, and where the local mass density is high, they acquire a mass, making their effects short range and unobservable. The effective coarse-grained description of chameleon theories, including careful considerations of averaging, has been derived in~\cite{Mota:2006fz}. Chameleonic vector fields, such as gauged $B-L$, have been proposed in~\cite{nelsonchamvec}. See~\cite{Hinterbichler:2010wu} for an attempt at realizing the chameleon in string compactifications.

Recently, two of us have proposed a third screening mechanism, called the symmetron~\cite{Hinterbichler:2010es}, based in part on earlier work~\cite{Olive:2007aj,Pietroni:2005pv}.
For this  mechanism to operate, the vacuum expectation value (VEV) of the scalar must depend on the local mass density. The VEV becomes large in
regions of low mass density, and small in regions of high mass density. In addition, the coupling of the scalar to matter is proportional to the VEV, so that
the scalar couples with gravitational strength in regions of low density, but is decoupled and screened in regions of high density.
 
\begin{figure}[t]
\centering
\subfloat[Effective potential in regions of high ambient density. The VEV is 0.]{\includegraphics[scale=.8]{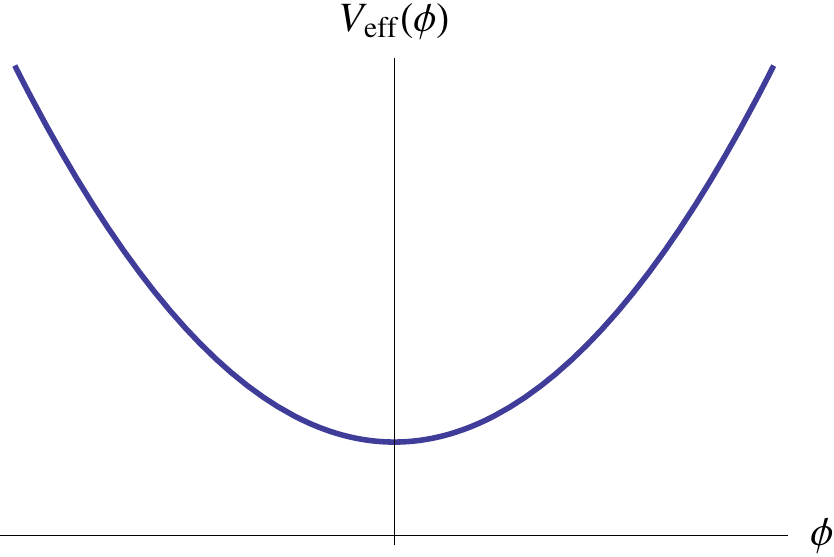}} \qquad
\subfloat[Effective potential in regions of low ambient density.]{\includegraphics[scale=.8]{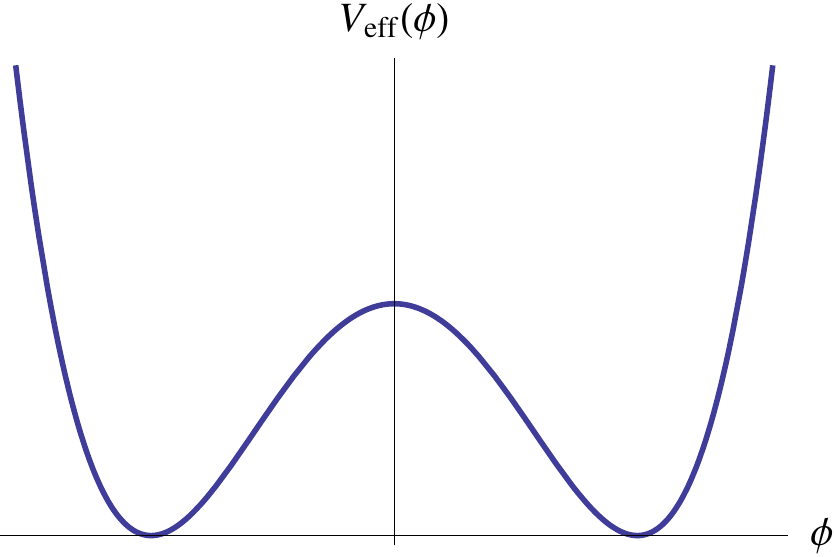}}
\caption{Schematic plots of the symmetron effective potential, illustrating the symmetry breaking phase transition.}
\label{sympot}
\end{figure}

This is achieved through the interplay of a symmetry-breaking potential~\cite{Hinterbichler:2010es}, 
\be
V(\phi) = -\frac{1}{2}\mu^2\phi^2 + \frac{1}{4}\lambda\phi^4\,, 
\label{quarticpot}
\ee
and a $\mathds{Z}_2$-invariant universal conformal coupling to the trace of the matter stress tensor, $\sim\phi^2 T^\mu_{\;\mu}/M^2$, so a local matter density contributes to the effective mass of the scalar.
In vacuum, the scalar acquires a VEV $|\phi| = \phi_0 \equiv \mu/\sqrt{\lambda}$, which spontaneously breaks
the reflection symmetry of the lagrangian $\phi\rightarrow -\phi$.  In the presence of sufficiently high ambient density, on the other hand,
the potential does not break the symmetry and the scalar is trapped near $\phi =0$. This is shown in Figure~\ref{sympot}. In addition, $\delta\phi$ fluctuations couple
to matter as $({\phi_{\rm VEV}}/M^2)\delta\phi\; \rho$. Hence, symmetron perturbations are weakly coupled in high
density backgrounds and relatively more strongly coupled in low density backgrounds.

The symmetron naturally takes the form of an effective field theory. The potential comprises the most general renormalizable terms invariant under
the $\mathds{Z}_2$ symmetry $\phi\rightarrow -\phi$. The coupling to matter is the leading such coupling compatible with the symmetry.  It is non-renormalizable, suppressed by the mass scale $M$, thus the symmetron is an effective theory with cutoff $M$. Remarkably, tests of gravity constrain this cutoff to be around the GUT scale~\cite{Hinterbichler:2010es}, so any GUT theory with a low energy scalar might be expected to yield a symmetron-type lagrangian at low energies. From this point of view, symmetron theories are more natural-looking than chameleon models. As with all non-supersymmetric scalars, however, the coupling to matter generates large quantum corrections to the mass which must be fine-tuned away. 

Symmetron theories predict a host of observational predictions, some of which are distinguishable from other screening mechanisms.
Although the local environment is dense, and therefore one in which the symmetron-mediated force is weak, the symmetron nevertheless
leads to small deviations from General Relativity in the solar system. As reviewed in Section~\ref{tests}, the predicted signals for Lunar Laser Ranging and Mercury's 
perihelion precession are just below current bounds and within reach of next-generation experiments~\cite{Hinterbichler:2010es}. 
On the other hand, the signal from binary pulsars is much weaker and should not be observed, since
both neutron stars are screened. This makes the symmetron distinguishable from over-the-counter scalar-tensor theories, such as
Brans-Dicke (BD) theories, for which solar system and binary pulsar signals are comparable. The symmetron solar system predictions
are also distinguishable from chameleon and Vainshtein screening~\cite{Hinterbichler:2010es}. 

Symmetrons, like chameleons, predict effective macroscopic violations of the equivalence principle between large (screened) and 
small (unscreened) bodies. Dwarf galaxies in low-density environments, in particular, offer a fertile playground for testing
these ideas~\cite{Hui:2009kc,Jain:2011ji}. The stars are oblivious to the symmetron whereas the hydrogen gas experiences an additional
force, resulting in an enhanced rotational velocity for the gas~\cite{Hui:2009kc}. Meanwhile, the infall motion of a dwarf galaxy can lead to
a segregation of the stellar disk from the dark matter and the hydrogen gas~\cite{Jain:2011ji}.

%% FIGURE
%% END FIGURE

We begin with an extensive overview of symmetron physics in Section~\ref{symrev}, reviewing
the spherically-symmetric solution (Section~\ref{sphsol}) and constraints from solar system tests of
gravity (Section~\ref{tests}). Section~\ref{jordan}, in particular, offers a novel perspective 
of the symmetron in terms of Jordan-frame variables. In this frame, the action describes a Brans-Dicke
type scalar-tensor theory, with a field-dependent Brans-Dicke parameter. The screening mechanism
is understood as the Brans-Dicke parameter becoming large in regions of high-density, corresponding
to the decoupling of the scalar. In this respect, the symmetron is qualitatively similar to the
Vainshtein mechanism, though its non-linearities stem from potentials rather than derivative self-couplings.

The remainder of the paper focuses on the cosmological evolution in the presence of a symmetron field.
Specifically, we are interested in checking whether a symmetron leads to an acceptable cosmology and, more interestingly, whether it can act as dark energy
driving cosmic acceleration. We derive the evolution of the scalar field from the inflationary epoch until the present.
The dependence of the symmetron effective potential on the matter density has implications for the cosmological evolution
of the scalar field. Since the matter density redshifts in time, the effective potential is time-dependent and results in a phase
transition when the matter density falls below a critical value. We will choose parameters such that the phase transition occurs
in the recent past ($z_{\rm tran} \,\lsim \, 1$). 

A key question is whether the evolution allows the scalar field to reach the symmetry-breaking vacuum by the present epoch, as assumed
in the analysis of solar system tests and other phenomenological studies. The short answer is that the coupling to matter efficiently
drives the field towards the symmetry-restoring point, so that for a
broad range of initial conditions it reaches $\phi\simeq 0$ well before the phase transition. How the field makes it there, however, is an interesting story, as summarized below.

In Section~\ref{cosmoevol} we describe the symmetron evolution during the radiation- and matter-dominated eras of standard
big bang cosmology.  The effective mass squared of the symmetron due to its coupling to matter fields is
$m_{\rm eff}^2 \sim T^\mu_{\;\mu}/M^2 \sim (1-3w)\rho/M^2$, where $w$ is the equation of state. During the radiation-dominated
era, $T^\mu_{\;\mu}$ is dominated by the (subdominant) non-relativistic component, hence
$m_{\rm eff}^2 \sim \Omega_{\rm m} H^2M_{\rm Pl}^2/M^2$. At early times, therefore, $m_{\rm eff}^2 \ll H^2$, and the symmetron 
remains essentially frozen at its initial value, denoted by $\phi_{\rm rad-i}$. As the universe cools and $\Omega_{\rm m}$ increases, eventually
$m_{\rm eff}^2 \simeq H^2$, and the symmetron starts rolling and undergoes damped oscillations around $\phi = 0$. Since $M \ll M_{\rm Pl}$,
this rolling phase is triggered deep in the radiation-dominated epoch, prior to Big Bang Nucleosynthesis (BBN). From this moment onwards,
the field amplitude decreases by a factor of $(M/M_{\rm Pl})^{3/2}$ until matter-radiation equality. The damped oscillations continue
during the matter-dominated era, and the field amplitude decreases by an additional factor of $10^{-3}$ until the present epoch.
Thus, for a broad range of initial field values, the symmetron ends up close to the origin by the onset of the phase transition.

In Section~\ref{phasetran}, we follow the evolution of the symmetron through the phase transition. For the quartic potential~(\ref{quarticpot}),
the symmetron nicely tracks the effective minimum and settles to the symmetry-breaking vacuum by the present time. However, 
the energy scale of the potential is too small to drive cosmic acceleration, and the symmetron backreaction is negligible. Hence we must
add a suitable cosmological constant, thereby making the expansion history indistinguishable from $\Lambda$CDM cosmology.

We also consider generalized, non-renormalizable forms of the potential, with scales chosen to have a greater impact on late-time cosmology: $i)$ the potential energy difference between maximum and minima is of order $H_0^2M_{\rm Pl}^2$, as desired to act as dark energy; and $ii)$ the mass of small fluctuations around the minima is of order $H_0$, such that the symmetron
can impact the growth of structure on the largest observable scales. The resulting potential is sharply peaked at $\phi = 0$, and displays very shallow minima. Because of this asymmetric form,
shortly after the onset of the phase transition the symmetron overshoots the minimum and reaches a value of order $M_{\rm Pl}$. As a result, the field does not converge to the symmetry-breaking vacuum by the present time, which is problematic for tests of gravity. Meanwhile, the symmetron energy density quickly becomes irrelevant after the transition. One way to circumvent these problems
is by requiring the phase transition to occur in the very recent past ($z_{\rm tran} \ll 1$), but this requires a delicate choice of parameters.

In Section~\ref{inf} we go back to the early universe and describe the evolution of the symmetron during a period of inflation and subsequent reheating.
Since the effective mass $m_{\rm eff}^2 \simeq 12 H^2_{\rm inf}M_{\rm Pl}^2/M^2$ is $\gg H^2_{\rm inf}$ during inflation,
the scalar undergoes rapid oscillations with exponentially decaying amplitude, thanks to Hubble damping. The end result is that a field
initially displaced from $\phi = 0$ ends up exponentially close to this point. At the end of inflation, however, as the inflaton oscillates around its minimum, its equation of state
also oscillates between $w = -1$ and $w = 1$. In turn, $m_{\rm eff}^2$ oscillates between negative and positive values,
which can result in amplification. 

With a simple parametrization of $w(t)$, we find that the symmetron evolution can be described
at early times by a Mathieu equation, with solution given by the Mathieu cosine function. As usual with Mathieu functions,
the stability behavior depends on parameter choices, in this case  the inflation mass (setting the rate of oscillations in $m_{\rm eff}^2$), its decay rate (setting the
duration of the unstable phase) and the effective symmetron mass during inflation. We place a constraint on these parameters by demanding that the symmetron
displacement be within effective field theory bounds.  We summarize our results and discuss future directions in Section \ref{conclsec}.

\section{Review of the Symmetron}
\label{symrev}

The symmetron is a special case of a general scalar-tensor theory,
\be
\label{mainlag}
S=\int {\rm d}^4x\ \sqrt{-g}\left[ {M_{\rm Pl}^2\over 2}R-\half g^{\mu\nu}\partial_\mu \phi\partial_\nu\phi-V(\phi)\right]+ \int {\rm d}^4x\ \sqrt{-\tilde{g}}{\cal L}_{\rm m}\left(\psi,\tilde{g}_{\mu\nu}\right)\,,
\ee
where the metric has mostly-plus signature, and ${\cal L}_{\rm m}$ is the lagrangian for the matter fields, denoted collectively by $\psi$.
The matter fields couple minimally to the Jordan frame metric, 
\be 
\tilde{g}_{\mu\nu}\equiv A^2(\phi) g_{\mu\nu}\,,
\ee
related to the Einstein frame metric $g_{\mu\nu}$ by a positive function $A(\phi)$. In particular, $\phi$ couples universally to all matter fields, hence the weak
equivalence principle holds. See \cite{lamEP} for a discussion of quantum corrections to universal coupling.

The scalar field equation of motion is given by
\be 
\square\phi = V_{,\phi} - A^3(\phi) A_{,\phi}(\phi) \tilde{T} \,,
\label{scalareqm}
\ee
where $\tilde{T} = \tilde{g}^{\mu\nu}\tilde{T}_{\mu\nu}$ is the trace of the the matter stress-energy tensor, $\tilde{T}_{\mu\nu} \equiv-(2/\sqrt{-\tilde{g}}) \delta{\cal L}_{\rm m}/ \delta \tilde{g}^{\mu\nu}$,
covariantly conserved with respect to $\tilde{g}_{\mu\nu}$: $\tilde{\nabla}_\mu  \tilde{T}^{\mu}_{\ \nu}=0$.
Meanwhile, the Einstein equations are
\be
M_{\rm Pl}^2 G_{\mu\nu} = T_{\mu\nu}^\phi+A^2(\phi)\tilde{T}_{\mu\nu}\,.
\label{metriceqnm}
\ee
Note that because $\phi$ couples conformally to matter, its stress-energy tensor,
\be
T_{\mu\nu}^\phi=\partial_\mu \phi \partial_\nu \phi-\frac{1}{2}g_{\mu\nu}(\partial \phi)^2 -g_{\mu\nu}V(\phi)\,,
\label{Tphi}
\ee
is of course not covariantly conserved: $\nabla^\mu T_{\mu\nu}^\phi \neq 0$.

The form of the functions $A(\phi)$ and $V(\phi)$ is crucial to the operation of the symmetron mechanism.  These functions are assumed symmetric under 
$\phi\rightarrow -\phi$ and are such that the effective  symmetry breaking potential (\ref{effectivep}) has a zero VEV for large $\rho$ and a large VEV for small $\rho$.
In addition, the function $A(\phi)$ should be such that the coupling of scalar fluctuations to matter is proportional to the VEV.

The simplest symmetron theory, considered in~\cite{Hinterbichler:2010es}, is that of a quartic potential~(\ref{quarticpot}) and quadratic coupling:
\be 
A(\phi)=1+{1\over 2M^2}\phi^2 + {\cal O}\left(\frac{\phi^4}{M^4}\right)\,.
\label{ourpotential} 
\ee
The potential $V(\phi)$ comprises the most general renormalizable form invariant under the $\mathds{Z}_2$ symmetry $\phi\rightarrow -\phi$.
The coupling to matter $\sim \phi^2/M^2$ is the leading such coupling compatible with the symmetry. However, as we will see shortly, once we impose
constraints from tests of gravity, the energy scale in this potential is too small to act as dark energy. In Section~\ref{generalpot},
we will consider generalizations that have a greater impact on cosmic acceleration.

The model involves two mass scales, $\mu$ and $M$, and one positive dimensionless coupling $\lambda$.  
The mass term is tachyonic, so that the $\mathbb{Z}_2$ symmetry $\phi\rightarrow -\phi$
is spontaneously broken. With this canonical choice, we will see shortly that the relevant field range
is $\phi \ll M$, such that any ${\cal O}(\phi^4/M^4)$ terms in $A(\phi)$ can be consistently neglected.
For the case of non-relativistic matter, which is relevant for most applications including local tests of
gravity, $\tilde{T} \simeq -\tilde{\rho}$. Expressing~(\ref{scalareqm}) in terms of $\rho = A^3(\phi)\tilde{\rho}$, which is 
$\phi$-independent, the effective potential is, up to an irrelevant constant,
\be
V_{\rm eff}(\phi)={1\over 2}\left({\rho\over M^2}-\mu^2\right)\phi^2+{1\over 4}\lambda\phi^4\,.
\ee
Whether the quadratic term is negative or not, and hence whether the $\mathbb{Z}_2$ symmetry is spontaneously broken or not,
depends on the local matter density. 

The screening mechanism works roughly as follows: in vacuum or in large voids, where $\rho\simeq 0$, 
the potential breaks reflection symmetry spontaneously, and the scalar acquires a VEV $|\phi| = \phi_0\equiv \mu/\sqrt\lambda$;
in regions of high density, such that  $\rho > M^2\mu^2$, the effective potential no longer breaks the symmetry,
and the VEV goes to zero. Meanwhile, to lowest order the symmetron-matter coupling is $\sim\rho \phi^2/M^2$.
Fluctuations $\delta\phi$ around the local background value $\phi_{\rm VEV}$, which would be detected by local
experiments, couple to density as
\be 
\sim{\phi_{\rm VEV}\over M^2}\delta\phi \ \rho\,.
\label{coupling}
\ee
In particular, the coupling is proportional to the local VEV.  In high-density environments where the symmetry is restored,
the VEV should be near zero and fluctuations of $\phi$ do not couple to matter. In less dense environments,
where $\rho < M^2\mu^2$ and the symmetry is broken, the coupling turns on. 

We will be interested in the case where the field becomes tachyonic around current cosmic density, 
\be
H_0^2M_{\rm Pl}^2 \sim \mu^2M^2\,,
\label{mucrit}
\ee
so that the phase transition occurs around the onset of cosmic acceleration.
This fixes $\mu$ in terms of $M$, and hence the mass $m_0$ of small fluctuations around the symmetry-breaking vacuum:
\be
m_0= \sqrt{2} \mu \sim \frac{M_{\rm Pl}}{M} H_0\,.
\label{muvalue}
\ee
As reviewed in Section~\ref{tests}, constraints from local tests of gravity require $M \,\lsim\, 10^{-4}M_{\rm Pl}$. 
Hence the range $m_0^{-1}$ of the symmetron-mediated force in regions of low mass density is $\,\lsim\, 0.1\;{\rm Mpc}$
--- too heavy to act as a slowly-rolling quintessence field, but light enough to impact structure formation and have
interesting astrophysical implications. 

The symmetron-mediated force $F_\phi$ relative to the Newtonian force $F_{\rm N}$ between two test masses in vacuum is set by the symmetry-breaking value $\phi_0$:
\be
\frac{F_\phi}{F_{\rm N}} = 2M_{\rm Pl}^2 \left(\frac{{\rm d}\ln A}{{\rm d}\phi}\bigg\vert_{\phi_0} \right)^2 \simeq 2\left(\frac{\phi_0 M_{\rm Pl}}{M^2}\right)^2\,.
\label{forceratio}
\ee
If the scalar-mediated force is to be comparable to gravity in vacuum, then we
must impose $\phi_0/M^2 \sim 1/M_{\rm Pl}$, that is,
\be
\phi_0\equiv \frac{\mu}{\sqrt{\lambda}} =  g \frac{M^2}{M_{\rm Pl}}\,,
\label{vev}
\ee
where $g \sim {\cal O}(1)$. To be precise, it follows from~(\ref{forceratio}) that $g$ measures the strength of scalar force in vacuum relative to gravity:
$F_\phi = 2g^2F_{\rm N}$.

Combined with~(\ref{muvalue}) and the requirement $M\,\lsim\, 10^{-4}M_{\rm Pl}$,~(\ref{vev}) fixes the dimensionless
quartic coupling to be exponentially small:
\be
\lambda \sim \frac{M_{\rm Pl}^4H_0^2}{M^6} \,\gsim\, 10^{-96}\,. 
\label{lamvalue}
\ee
Note that since $M \ll M_{\rm Pl}$ from tests of gravity considerations,~(\ref{vev}) implies that $\phi \ll M$,
hence the entire field range of interest lies within the regime of the effective field theory.

Unfortunately, these requirements imply that the potential energy is too small to have significant backreaction and
drive cosmic acceleration. Indeed, using~(\ref{mucrit}) and~(\ref{vev}), the potential height difference
between $\phi = 0$ and $\phi =\phi_0$ is
\be
\Delta V = \frac{\mu^4}{4\lambda} \sim H_0^2 M^2 \ll H_0^2 M_{\rm Pl}^2\,.
\label{delV}
\ee
In Section~\ref{generalpot} we will consider generalized forms of the potential that have $\Delta V \sim H_0^2M_{\rm Pl}^2$ and are therefore
better candidates for dark energy applications. 

\subsection{Jordan-Frame Description}
\label{jordan}

Because the scalar is assumed to couple universally to matter, we can also understand the symmetron mechanism in Jordan frame. 
We will see that the coefficient of the symmetron kinetic term depends on the local value of the scalar field. This coefficient becomes
large in regions of high density, which effectively decouples the scalar field from matter. From this point of view, the symmetron mechanism
is qualitatively similar to Vainshtein screening, though it relies on a potential rather than derivative self couplings.

In terms of the Jordan-frame metric $\tilde{g}_{\mu\nu} = A^2(\phi) g_{\mu\nu}$, the action~(\ref{mainlag}) takes the form
\be
S = \int {\rm d}^4x\ \sqrt{-\tilde{g}}\left[ {M_{\rm Pl}^2\over 2}\Psi \tilde{R}-\frac{M_{\rm Pl}^2\omega(\Psi)}{2\Psi} \tilde{g}^{\mu\nu}\partial_\mu \Psi\partial_\nu\Psi-\Psi^2 V(\Psi)\right]
+ \int {\rm d}^4x\ \sqrt{-\tilde{g}}{\cal L}_{\rm m}\left(\psi,\tilde{g}_{\mu\nu}\right)\,,
\label{SJordan}
\ee
where $\Psi \equiv A^{-2}(\phi)$. Thus the symmetron behaves as a Brans-Dicke scalar field, with field-dependent Brans-Dicke parameter
\be
\omega(\phi) = \frac{1}{2}\left[ \frac{1}{2M_{\rm Pl}^2 \left({\rm d}\ln A/{\rm d}\phi\right)^2} - 3\right] \simeq  \frac{1}{2}\left[ \frac{1}{2}\left(\frac{M^2}{M_{\rm Pl}\phi}\right)^2 - 3\right]\,,
\ee
where in the last step we have substituted the quadratic form $A(\phi) \simeq 1 +  \phi^2/2M^2$ and used the fact that $A(\phi) \simeq 1$ for all field values of interest.

The nature of the screening mechanism in this frame is crystal clear. In regions of high density, where $\phi \simeq 0$, the Brans-Dicke parameter is large,
\be
\omega \simeq \left(\frac{M^2}{2M_{\rm Pl}\phi}\right)^2 \gg 1 \qquad ({\rm high}\; {\rm density}\; {\rm regions})\,,
\label{omhigh}
\ee
indicating that the scalar field decouples and General Relativity is recovered. In regions of low density, where $|\phi| \simeq \phi_0 = g M^2/M_{\rm Pl}$, the Brans-Dicke parameter is of
order unity,
\be
2 + 3\omega \simeq\frac{1}{2g^2}\qquad ({\rm low}\;\; {\rm density}\; {\rm regions})\,,
\ee
hence the scalar field generates order unity corrections to General Relativity.

\subsection{Spherically Symmetric Solution}
\label{sphsol}

To study the implications for tests of gravity, we are interested in the symmetron profile around astrophysical sources,
such as the Sun. For this purpose, we can safely work in the Newtonian limit, ignoring non-linear effects in gravity and
the back-reaction of the scalar field on the metric. Moreover, the source is approximated as static,
spherically symmetric and pressureless ($\tilde{T} \simeq -\tilde{\rho}$). It has radius $R$
and homogeneous mass density $\rho$, such that $\rho \gg \mu^2M^2$. For simplicity, we further assume the object lies
in vacuum. 

Written in terms of the density $\rho=A^{3}\tilde\rho$, which is conserved in Einstein frame, the scalar field equation~(\ref{scalareqm}) in spherical coordinates
reduces to
\be 
{{\rm d}^2\over {\rm d}r^2}\phi+{2\over r}{{\rm d}\over {\rm d}r}\phi=V_{,\phi}+ A_{,\phi}\rho\,.
\label{sphericalequation} 
\ee
Analogously to what was done in~\cite{Khoury:2003aq}, this radial field equation can be thought of as a fictional particle rolling in an inverted effective potential $-V_{\rm eff} (\phi)$,
subject to a ``friction'' term $(2/r){\rm d}\phi/{\rm d}r$. The boundary conditions are that the solution be smooth at the origin, and approach its symmetry-breaking value at infinity:
\be
\frac{{\rm d}}{{\rm d}r}\phi(0)=0\,;\ \ \ \ \phi(r\rightarrow \infty)=\phi_0\,.
\label{boundaryconditions}
\ee

First consider the solution inside the object. Since the objects of interest are much denser than the current critical density, $\rho \gg \mu^2M^2\sim H_0^2M_{\rm Pl}^2$,
the effective potential can be approximated as  $V_{\rm eff}(\phi) \simeq \rho\phi^2/2M^2$. The interior solution satisfying the first of~(\ref{boundaryconditions}) is
\be 
\phi_{\rm in}(r) \simeq C\, {R\over r}\sinh\left( \frac{\sqrt{\rho}}{M} r\right)  \qquad ({\rm for}\; r < R)\,.
\label{insol}
\ee
This involves one undetermined constant $C$ to be fixed shortly.

Next consider the exterior solution. It turns out that the potential is negligible, and the symmetron evolves as a free field, until it reaches the vicinity of the minimum. 
In the mechanical analogy, this corresponds to the fictional particle having a large velocity as it exits the object and begins its climb towards the maximum of
the inverted potential. Since the potential is irrelevant except near $\phi = \phi_0$, we can make the quadratic approximation $V_{\rm eff}(\phi) = m_0^2(\phi-\phi_0)^2/2$ for all $r > R$. The solution satisfying the second of (\ref{boundaryconditions}) is then given by
\be 
\phi_{\rm out}(r)=D\,{R\over r}e^{-m_0 (r-R)}+\phi_0 \qquad ({\rm for}\; r > R)\,,
\label{outsol}
\ee
with one undetermined constant $D$.

The coefficients $C$ and $D$ are fixed by matching the field and its radial derivatives at the interface
$r = R$. In doing the matching, we can safely assume that $m_0 R \ll 1$, since the symmetron Compton wavelength
$m_0^{-1}$ is cosmologically large ($ \,\lsim\, 0.1\; {\rm Mpc}$). It is convenient to express the result as
\bea
\nonumber
C &=&  \phi_0 \sqrt{\frac{\Delta R}{R}} {\rm sech}\left(\sqrt{\frac{R}{\Delta R}}\right) \\
D &=& -\phi_0 \left[1-\sqrt{\frac{\Delta R}{R}} {\rm tanh}\left(\sqrt{\frac{R}{\Delta R}}\right) \right]\,.
\label{coeffs}
\eea
where we have introduced the {\it thin-shell} factor,
\be
\frac{\Delta R}{R} \equiv \frac{M^2}{\rho R^2} = \frac{M^2}{6M_{\rm Pl}^2\Phi} =  \frac{\phi_0}{6g M_{\rm Pl}\Phi}\,.
\label{thinshell}
\ee
Here $\Phi \equiv \rho R^2/6M_{\rm Pl}^2$ denoting the gravitational potential of the source, and where in the last step
we have used~(\ref{vev}). As the notation suggests, $\Delta R/R$ will soon be interpreted as a thin shell factor for the solutions, in analogy with
the chameleon mechanism. In fact,~(\ref{thinshell}) precisely matches the chameleon thin shell expression,
with $\phi_0$ being interpreted as the ambient chameleon value. Hence one can immediately anticipate that symmetrons
and chameleons will have similar phenomenology, in particular for astrophysical tests~\cite{Hui:2009kc,Jain:2011ji}. 

Consider a test particle a distance $R \ll r \ll m_0^{-1}$ away from the object, such that $\phi\simeq \phi_0$ and ${\rm d}\ln A/{\rm d}\phi \simeq g/M_{\rm Pl}$.
Substituting the expression for $D$ into~(\ref{outsol}) and using~(\ref{thinshell}), the scalar force on this particle relative to gravity is
\be
\frac{F_{\phi}}{F_{\rm N}} = - \frac{g}{M_{\rm Pl}}\frac{{\rm d}\phi/{\rm d}r}{F_{\rm N}} = 6g^2 \frac{\Delta R}{R} \left[1-\sqrt{\frac{\Delta R}{R}} {\rm tanh}\left(\sqrt{\frac{R}{\Delta R}}\right) \right]\,.
\label{forcerationew}
\ee
Aside from the coefficient $g$, the magnitude of the scalar force is thus determined by a single parameter, $\Delta R/R$.

{\it Screened objects} have relatively large gravitational potential such that $\Delta R/R \ll 1$. In this case the tanh term can be neglected and~(\ref{forcerationew}) reduces to
\be
\frac{F_{\phi}}{F_{\rm N}}\bigg\vert_{\rm screened} \simeq 6g^2 \frac{\Delta R}{R}\ll 1\,.
\ee
The scalar force between a screened object and a test particle is therefore suppressed by a thin-shell factor $\Delta R/R \ll 1$.
This can be understood as an analogue of the thin-shell effect of chameleon models~\cite{Khoury:2003aq,Khoury:2003rn}.
It is clear from the interior profile, 
\be
\phi_{\rm in}(r) \simeq \phi_0 \sqrt{\frac{\Delta R}{R}} \frac{R}{r} \frac{\sinh\left(\sqrt{\frac{R}{\Delta R}}\frac{r}{R}\right)}{\cosh\left(\sqrt{\frac{R}{\Delta R}}\right)}\,,
\label{phiin}
\ee
that the field is exponentially suppressed compared to $\phi_0$ everywhere inside the object, except within a thin-shell $\Delta R$ beneath the surface.
The symmetron is thus weakly coupled to the core of the object, hence its exterior profile is dominated by the thin shell contribution.

{\it Unscreened objects}, on the other hand, have relatively small gravitational potential such that $\Delta R/R$ in~(\ref{thinshell}) is formally $\gg 1$.
In this regime, we can Taylor expand the tanh and~(\ref{forcerationew}) simplifies to
\be
\frac{F_{\phi}}{F_{\rm N}} \simeq 2g^2\,.
\ee
There is no thin shell in this case, and the result is consistent with the scalar force between two test masses
given by~(\ref{forceratio}) and~(\ref{vev}). Indeed, we see from~(\ref{phiin}) that $\phi\simeq \phi_0$ everywhere inside the object,
hence the symmetron couples with gravitational strength to the entire source.

\subsection{Constraints from Tests of Gravity}
\label{tests}

For the theory to be phenomenologically viable, it must of course satisfy all constraints from tests of gravity.
Here we review and expand on the original analysis of~\cite{Hinterbichler:2010es}. Because the symmetron has a
long Compton wavelength in all situations of interest, tests of the inverse-square-law are trivially satisfied. And
because its coupling to matter is universal, the weak equivalence principle is also satisfied. In fact, as shown in Section~\ref{jordan},
the symmetron can be understood as a Brans-Dicke scalar field with an effectively density-dependent Brans-Dicke parameter.
Hence the relevant tests are the same that apply to standard Brans-Dicke theories, namely post-Newtonian tests in the
solar system and binary pulsar observations.

What matters for solar system tests is the local field value, since this determines the coupling of the symmetron to matter.
At a generic point in the solar system, this is determined by the symmetron profile interior to the galaxy. Clearly a necessary
condition to satisfy local tests is that the Milky Way galaxy be screened, $\Delta R_{\rm G}/R_{\rm G} \ll 1$, in which case
the local value is given by~(\ref{phiin})
\be
\phi_{\rm G} \simeq \phi_0 \sqrt{\frac{\Delta R_{\rm G}}{R_{\rm G}}} \frac{R_{\rm G}}{R_{\rm us}} \frac{\sinh\left(\sqrt{\frac{R_{\rm G}}{\Delta R_{\rm G}}}\frac{R_{\rm us}}{R_{\rm G}}\right)}{\cosh\left(\sqrt{\frac{R_{\rm G}}{\Delta R_{\rm G}}}\right)}\,,
\label{phiinG}
\ee
where $R_G\sim 100$~kpc is the Milky Way radius, and $R_{\rm us}\sim 10$~kpc is our distance from the galactic center.
Recall that for the quartic potential considered here, $\phi_0 = g M^2/M_{\rm Pl}$. We will see that post-Newtonian constraints
in the solar system translate to a bound on $\Delta R_{\rm G}/R_{\rm G}$, and hence on $M$.

General scalar-tensor theories of the form~(\ref{SJordan}) have two non-vanishing post-Newtonian parameters, $\beta$ and $\gamma$,
defined in terms of the Jordan-frame metric as
\bea
\nonumber
\tilde{g}_{00} &=& - \left(1 + 2\Phi_{\rm J} + 2\beta \Phi_{\rm J}^2\right)\;; \\
\tilde{g}_{ij} &=& \left(1 - 2\Phi_{\rm J}\gamma\right)\delta_{ij} \,.
\eea
In terms of $\omega(\phi)$, these parameters are given by~\cite{Will:2005va}
\bea
\nonumber
\gamma &=& \frac{1 + \omega(\phi)}{2 + \omega(\phi)}  \;; \\
\beta &=& 1 + \frac{1}{\left(3+ 2\omega\right)^2 \left(4 + 2\omega\right)} \frac{{\rm d}\omega}{{\rm d}(A^{-2})}\,.
\eea
From~(\ref{omhigh}), the local value of these parameters satisfy
\bea
\nonumber
|\gamma - 1|_{\rm G} &\simeq& \left(\frac{2M_{\rm Pl}\phi_{\rm G}}{M^2}\right)^2 = 4g^2\left(\frac{\phi_{\rm G}}{\phi_0}\right)^2\;; \\
|\beta - 1 |_{\rm G} &\simeq & 2g^2 \frac{M_{\rm Pl}^2}{M^2} \left(\frac{\phi_{\rm G}}{\phi_0}\right)^2\,.
\label{PPNdev}
\eea
The tightest constraint on $\gamma$ comes from time-delay measurements with the Cassini spacecraft~\cite{Bertotti:2003rm}: $|\gamma - 1| \,\lsim\, 10^{-5}$.
The most stringent bound on $\beta$ comes from the Nordvedt effect, which describes the difference in free-fall acceleration of the Moon and the
Earth towards the Sun due to scalar-induced differences in their gravitational binding energy. Lunar Laser Ranging observations give the constraint
$|\beta - 1 |\,\lsim\, 10^{-4}$~\cite{Will:2005va}. 

Although the experimental bound on $\beta$ is weaker, it actually yields the tightest constraint on our model parameters, since the predicted deviation in~(\ref{PPNdev})
is $\sim M_{\rm Pl}^2/M^2$ larger than $\gamma$. Assuming $g\sim {\cal O}(1)$ and using the fact that $\Phi_{\rm G}  \simeq 10^{-6}$ for the Milky Way galaxy, we
find that $|\beta - 1 |\,\lsim\, 10^{-4}$ is satisfied for
\be
\frac{\Delta R_{\rm G}}{R_{\rm G}} \,\lsim\, 6\times 10^{-3}\,.
\label{DelRG}
\ee
In terms of the coupling mass scale $M$, this translates to
\be
M \,\lsim\, 10^{-4}\; M_{\rm Pl}\,.
\label{Mbound}
\ee
In turn, this corresponds to $|\gamma - 1|_{\rm G} \sim 10^{-10}$, which lies well below the current limit. Intriguingly, the upper bound on $M$ is near the GUT scale.
Note that if $M$ nearly saturates this bound, then objects with $\Phi \,\gsim\, 10^{-8}$, such as the Sun ($\Phi_\odot\sim 10^{-6}$) will be screened
but those with $\Phi \,\lsim\, 10^{-8}$, such as the Earth ($\Phi_\oplus\sim 10^{-9}$), will not. 

If the symmetron is a low energy scalar associated with GUT-scale physics, such that $M$ nearly saturates~(\ref{Mbound}), then near-future experiments probing
post-Newtonian corrections can potentially detect symmetron effects. The APOLLO observatory, in particular, should improve Lunar Laser Ranging constraints by
an order of magnitude~\cite{APOLLO}. As we have seen, the predicted signal for time-delay and light-deflection are much more suppressed.

Meanwhile, the constraints from binary pulsars are trivially satisfied, since both the neutron star and its companion are screened.
The force between these bodies is therefore suppressed by two thin-shell factors:
\be
\frac{F_\phi}{F_{\rm N}} = \left(\frac{\Delta R}{R}\right)_{\rm pulsar}\times \left(\frac{\Delta R}{R}\right)_{\rm companion} \sim 10^{-5} \left(\frac{\Delta R_{\rm G}}{R_{\rm G}}\right)^2 \,\lsim\, 10^{-10}\,,
\ee
where we have estimated $\Phi_{\rm pulsar} \sim 0.1 \sim 10^5\Phi_{\rm G} $ and $\Phi_{\rm companion}\sim 10^{-6} \sim \Phi_{\rm G}$, and the last step follows from~(\ref{DelRG}).
This is well below the current pulsar constraints on Brans-Dicke scalar-tensor theories, which translate to $F_\phi/F_{\rm N}\, \lsim \, 10^{-4}$~\cite{Will:2005va}. 

\section{Cosmological Evolution}
\label{cosmoevol}

Central to the phenomenological viability of symmetron theories is whether the resulting expansion and growth histories are
consistent with observations. In the remainder of the paper, we will study the cosmological evolution of the symmetron field, 
from the inflationary epoch until today. In this Section, we begin by setting up the equations and solving them for
the standard radiation and matter dominated eras. Given the scales involved, the scalar potential is completely negligible
at early times, and the symmetron evolution is determined by the interplay of Hubble friction and its direct coupling to matter.

A key question is whether the evolution allows the scalar field to reach the symmetry-breaking vacuum by the present epoch, as assumed
in the analysis of local tests discussed in Section~\ref{tests}. We will see that the coupling to matter is efficient in bringing
the field close to $\phi = 0$ by the onset of the phase transition ($z \sim 1$), for a broad range of initial conditions.

\subsection{Preliminaries}

Consider the homogeneous evolution of the scalar field, $\phi = \phi(t)$, in a Friedmann-Robertson-Walker (FRW) background:
${\rm d} s^2_{\rm E} = -{\rm d}t^2  + a^2(t){\rm d}\vec{x}^2$. We assume that inflation takes place at early times, so that the universe
is nearly spatially-flat. The above line element is in Einstein frame, hence the label ``E". Throughout the analysis, dots will represent
derivatives with respect to (Einstein-frame) cosmic time $t$.

The matter content is modeled as a set of non-interacting perfect fluids indexed by $i$, each with constant equation of state $w_i$.
The trace of the stress tensor therefore becomes $\tilde{T}=\sum_i (-1+3 w_i)\tilde{\rho}_i$. Substituting this into~(\ref{scalareqm}),
the symmetron equation of motion reduces to
\be 
\ddot\phi+3H\dot\phi+V_{,\phi}  + A^3(\phi) A_{,\phi} \sum_i(1-3w_i)\tilde{\rho}_i=0\,. 
\label{scalarfrw}
\ee
By assumption, each matter component is separately conserved:
\be
\tilde{\rho}_i \sim a_{\rm J}^{-3(1 + w_i)} = A^{-3(1+w_i)} a^{-3(1+w_i)}\,,
\ee
where $a_{\rm J} = A(\phi)a$ is the Jordan-frame scale factor. To make the $\phi$-dependence explicit in~(\ref{scalarfrw}),
it is convenient to define a rescaled energy density,
\be 
\rho_i\equiv A^{3(1+w_i)}\tilde\rho_i \,,
\label{rhorhot}
\ee
such that $\rho_i \sim a^{-3(1 + w_i)}$ satisfies the usual conservation law in Einstein frame. Note that $\rho_i$ does not represent the physical density
in Einstein frame --- it is a mathematical construct, which can be treated as independent of $\phi$ when integrating~(\ref{scalarfrw}). 

In terms of $\rho_i$, the symmetron equation of motion becomes
\be
\ddot\phi+3H\dot\phi+V_{,\phi} +  \sum_i{A_{,\phi} \over A^{3w_i}}(1- 3w_i){\rho_i}=0\,.
\label{scalarfrwfinal}
\ee
For regions of nearly constant $\rho_i$, we can think of this as a scalar field moving in an effective potential 
\be 
V_{\rm eff}(\phi)=V(\phi)+\sum_i A^{1-3w_i}(\phi)\rho_i\,. 
\label{effectivep}
\ee
As the universe expands and the densities $\rho_i$ redshift, this effective potential correspondingly evolves in time. Note that the difference between $\tilde{\rho}_i$ and $\rho_i$ is negligible for most of the history of the universe, since $A(\phi) \simeq 1$ until the phase transition.

Meanwhile, the Einstein-frame Friedmann equation, obtained from the $(0,0)$ component of~(\ref{metriceqnm}), is given by
\be
3M_{\rm Pl}^2H^2=\half \dot\phi^2+V(\phi)+A^4(\phi) \sum_i\tilde\rho_i \,.
\ee
In terms of the $\phi$-independent density $\rho_i$, this becomes
\be
3M_{\rm Pl}^2H^2=\half \dot\phi^2+V(\phi)+\sum_i A^{1-3w_i}(\phi)\rho_i =\half \dot\phi^2+V_{\rm eff}(\phi)  \,.
\label{friedeq}
\ee
In summary, the independent cosmological equations are the scalar field equation~(\ref{scalarfrwfinal}), the Friedmann equation~(\ref{friedeq}) and
the conservation laws $\rho_i\sim a^{-3(1+w_i)}$. In the remainder of this Section, we will study the evolution of the scalar field during the standard
radiation- and matter-dominated eras. The most recent history, including the phase transition and the onset of cosmic acceleration
will be discussed in Section~\ref{phasetran}. 

\subsection{Radiation-Dominated Era}
\label{RDera}

We imagine that after inflation the symmetron ends up at rest at some field value $\phi_{\rm rad-i}$. Later on in Section~\ref{inf} we will estimate $\phi_{\rm rad-i}$ by studying the inflationary and reheating dynamics in detail, but for the time being let us consider it as a free parameter. To simplify the analysis, we assume that $\phi_{\rm rad-i} \ll M$, so that higher-order corrections in~(\ref{ourpotential}) for the coupling function are negligible:
\be
A(\phi) \simeq 1 + \frac{\phi^2}{2M^2}\,. 
\label{Aphisimplify}
\ee
Moreover, the self-interaction potential $V(\phi)$ can be neglected until the late-time phase transition. Indeed, given~(\ref{muvalue}) and~(\ref{lamvalue}) the quartic
potential satisfies $V(\phi) \simeq V(0) \sim \mu^4/\sqrt{\lambda} \ll H_0^2M_{\rm Pl}^2$. The symmetron kinetic energy is also negligible, as argued shortly.
Hence, for most of the evolution the symmetron dynamics are governed by a tug-of-war between the coupling to matter driving the
field and Hubble friction slowing it down.

Since $A(\phi)\simeq 1$, the Friedmann equation~(\ref{friedeq}) in the radiation-dominated epoch takes the standard form
\be
3H^2M_{\rm Pl}^2 = \rho_{\rm r}(a)\,,
\ee
where $\rho_{\rm r} \sim a^{-4}$. Meanwhile, since the radiation stress energy has negligible trace, it drops out of the scalar equation of motion, and~(\ref{scalarfrwfinal}) reduces to
\be
\ddot\phi + 3H\dot\phi +  \rho_\mathrm{m} (a) \frac{\phi}{M^2}=0\,,
\label{scalarRD}
\ee
where $\rho_{\rm m} \sim a^{-3}$. Note that we have neglected the potential term $V_{,\phi}$, by the argument below~(\ref{Aphisimplify}).
The physics of this equation is clear --- it describes a simple harmonic oscillator subject to friction with a weakening spring constant
(weakening since $\rho_{\rm m}$ redshifts in time). At very early times, when $H^2 \gg \rho_{\rm m}/M^2$, the oscillator is overdamped,
and the field remains stuck at $\phi \simeq \phi_{\rm rad-i}$. At late times, when $H^2 \ll \rho_{\rm m}/M^2$, the oscillator is underdamped,
and the symmetron oscillates around its minimum. In other words, we expect the field to remain frozen at its initial value until $a = a_{\rm unfreeze}$,
where $a_{\rm unfreeze}$ satisfies
\be
1 = \frac{\rho_{\rm m}(a)}{H^2M^2} = \frac{3M_{\rm Pl}^2}{M^2} \frac{a_{\rm unfreeze}}{a_{\rm eq}}\,.
\ee
Here $a_{\rm eq}$ is the scale factor at matter-radiation equality. That is,
\be
a_{\rm unfreeze} = a_{\rm eq} \frac{M^2}{3M_{\rm Pl}^2}\,.
\label{aunfreeze}
\ee
Since $M^2 \ll M_{\rm Pl}^2$, the scalar field begins to roll well before matter-radiation equality. For instance, with $M = 10^{-4}\; M_{\rm Pl}$,
this occurs at $z \simeq 10^{12}$, {\it i.e.}, well before BBN.

These considerations are borne out by the explicit solution. Defining
\be
s \equiv \frac{2\sqrt{3}M_{\rm Pl}}{M}\left(\frac{a}{a_{\rm eq}}\right)^{1/2}\,,
\ee
the scalar equation of motion~(\ref{scalarRD}) simplifies to
\be
s^2\frac{{\rm d}^2\phi}{{\rm d}s^2} + 3s \frac{{\rm d}\phi}{{\rm d}s} + s^2\phi = 0\,.
\ee
The solution with $\phi \rightarrow \phi_{\rm rad-i}$ for $s\ll 1$ is given by a Bessel function:
\be
\phi = \frac{2\phi_{\rm rad-i}}{s}J_1(s)\,.
\label{besselsol}
\ee
As advocated, the field remains essentially constant until $s \sim 1$, which indeed
corresponds to $a \sim a_{\rm unfreeze}$. Subsequently, the field undergoes damped oscillations. At late times $s \gg 1$, in particular,~(\ref{besselsol}) implies
\be
\phi \sim \phi_{\rm rad-i} \frac{\cos s}{s^{3/2}} \sim \phi_{\rm rad-i}\left(\frac{M}{M_{\rm Pl}}\right)^{3/2} \left(\frac{a_{\rm eq}}{a}\right)^{3/4} \cos\left(\frac{2\sqrt{3}M_{\rm Pl}}{M}\left(\frac{a}{a_{\rm eq}}\right)^{1/2}\right) 
\,,
\label{RDsol}
\ee
up to an irrelevant phase. Thus, by matter-radiation equality the amplitude of the scalar field has dropped by a factor of
\be
\frac{|\phi_{\rm eq}|}{|\phi_{\rm rad-i}|} \sim \left(\frac{M}{M_{\rm Pl}}\right)^{3/2}\,.
\label{RDdecay}
\ee
With $M = 10^{-4}M_{\rm Pl}$, for instance, this corresponds to a suppression factor of $10^{-6}$.

We are now in a position to check that the symmetron kinetic energy is indeed a negligible contribution
in the Friedmann equation, as assumed throughout. From the properties of Bessel functions, and since $H\sim a^{-2}$ in the radiation-dominated epoch,
we have
\be
\frac{|\dot{\phi}|}{H} = \frac{s}{2} \left\vert \frac{{\rm d}\phi}{{\rm d}s}\right\vert \,\lsim\, \phi_{\rm rad-i}\ll M_{\rm Pl}\,.
\ee
It follows that $\dot{\phi}^2 \ll H^2 M_{\rm Pl}^2$, which confirms our approximation. Physically, this can also be understood
from~(\ref{scalarRD}), describing a damped oscillator with potential energy $\rho_{\rm m}\phi^2/2M^2$. Once the field
starts oscillating, the friction term can be ignored over the course of a few oscillations. The virial theorem then implies
$\langle \dot{\phi}^2\rangle = \rho_{\rm m}\langle \phi^2\rangle/M^2 < \rho_{\rm m} \phi_{\rm rad-i}^2/M^2 \ll \rho_{\rm m}$,
where $\phi_{\rm rad-i} \ll M$ by assumption. The symmetron kinetic energy is therefore subdominant to the matter component.

\subsection{Matter-Dominated Era}

In the matter-dominated era, $\rho_{\rm m} \simeq 3H^2M_{\rm Pl}^2$, the scalar equation of motion becomes
\be
\frac{{\rm d}^2\phi}{{\rm d}N^2} + \frac{3}{2}\frac{{\rm d}\phi}{{\rm d}N} + \frac{3M_{\rm Pl}^2}{M^2}\phi = 0\,,
\label{scalarMD}
\ee
where $N = \ln (a/a_{\rm eq})$ is the number of e-folds since matter-radiation equality. Up to an irrelevant phase,
the solution with initial condition $\phi = \phi_{\rm eq}$ at equality is
\be
\phi = \phi_{\rm eq} e^{-3N/4}\cos\left(\sqrt{\frac{3M_{\rm Pl}^2}{M^2} - \frac{9}{16}}N\right) = \phi_{\rm eq} \left(\frac{a_{\rm eq}}{a}\right)^{3/4}\cos\left(\sqrt{\frac{3M_{\rm Pl}^2}{M^2} - \frac{9}{16}}  \ln \left(\frac{a}{a_{\rm eq}}\right)  \right) \,.
\label{MDsol}
\ee
Again the field undergoes damped oscillations. By the onset of the phase transition ($z\simeq 1$), the amplitude has dropped by
\be
\frac{|\phi_0|}{|\phi_{\rm eq}|} \sim \left(\frac{2}{1+z_{\rm eq}}\right)^{3/4}  \sim 10^{-3}\,.
\label{MDdecay}
\ee

Interestingly, note from~(\ref{RDsol}) and~(\ref{MDsol}) that $\phi \sim a^{-3/4}$ both in the radiation- and matter-dominated eras.
Combining~(\ref{RDdecay}) and~(\ref{MDdecay}), the total decay in the field amplitude from the moment it starts to evolve deep
in the radiation-dominated era to the onset of the phase transition is
\be
\frac{|\phi_0|}{|\phi_{\rm rad-i}|} = 10^{-3} \left(\frac{M}{M_{\rm Pl}}\right)^{3/2}\,.
\ee

\begin{figure}[h]
\centering
\subfloat[]{\includegraphics[scale=1]{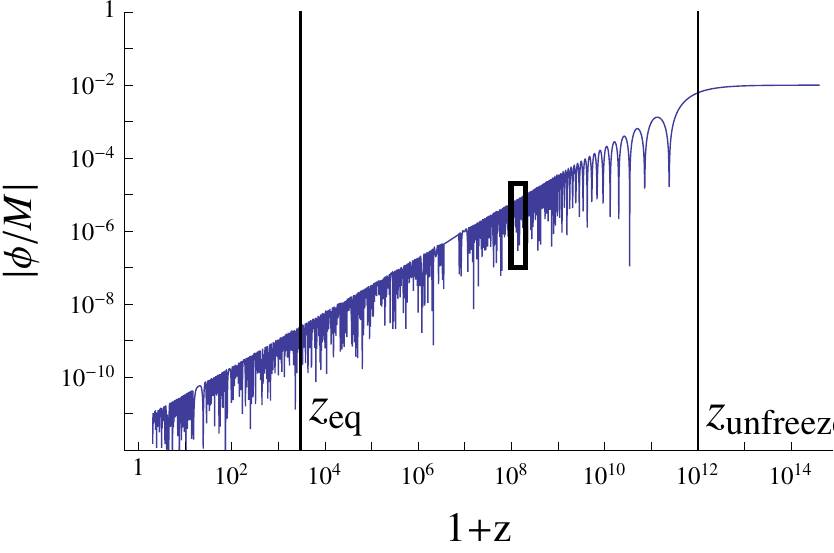}} \,
\subfloat[]{\includegraphics[scale=1]{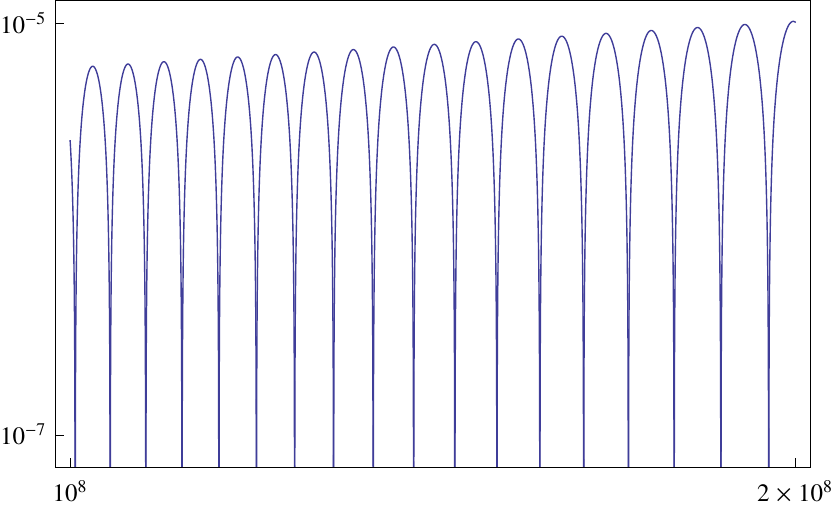}} 
\caption{Numerical solution of the symmmetron evolution as a function of redshift through the radiation- and matter-dominated eras, starting from $\phi_{\rm rad-i} = 10^{-2}M_{\rm Pl}$,
with $M = 10^{-4}M_{\rm Pl}$. Figure (a) confirms that the field is initially frozen because of Hubble friction, but eventually undergoes damped oscillations. See main text for details. 
Figure (b) is a zoom-in on the boxed region, highlighting that the oscillations are smooth.} 
\label{numevol}
\end{figure}

These results have been confirmed numerically. Figure~\ref{numevol}a) shows the results of integrating the combined scalar field equation~(\ref{scalarfrwfinal}) and Friedmann equation~(\ref{friedeq}) from $z = 10^{15}$ to $z = 1$, with $\phi_{\rm rad-i} = 10^{-2}M$ and $M = 10^{-4}M_{\rm Pl}$. For this purpose, we approximated $A(\phi)$ as the quadratic form~(\ref{Aphisimplify}) and chose $\mu = M_{\rm Pl}H_0/M$ and $\lambda = M_{\rm Pl}^4H_0^2/M^6$, as in~(\ref{muvalue}) and~(\ref{lamvalue}), respectively, even though $V(\phi)$ is irrelevant at these redshifts. As expected, the field remains stuck at its initial value until $z_{\rm unfreeze} \sim 10^{12}$, in agreement with~(\ref{aunfreeze}). The field subsequently undergoes damped oscillations, with an amplitude decaying as $(1+z)^{3/4}$. By matter-radiation equality, the field amplitude has decreased by a factor of $\sim 10^{-6}$, consistent with~(\ref{RDdecay}). During the matter-dominated era, the amplitude decays by an additional factor of $\sim 10^{-3}$, which agrees with~(\ref{MDdecay}). Figure~\ref{numevol}b) zooms in on the boxed region in Figure~\ref{numevol}a), highlighting the fact that the oscillations are completely smooth, despite appearing jagged in Figure~\ref{numevol}a).

%%% FIGURE
%% END FIGURE

\section{Phase Transition and Late-Time Evolution}
\label{phasetran}

In the previous Section, we have seen that starting from $\phi_{\rm rad-i} \,\lsim\, M$ deep in the radiation-dominated phase, the symmetron 
subsequently evolves towards the symmetry-restoring point, culminating to a value $\sim 10^{-3}(M/M_{\rm Pl})^{3/2}\phi_{\rm rad-i}$
by the onset of the phase transition.

In this Section, we describe the symmetron dynamics during this late-time phase transition. For the quartic potential~(\ref{quarticpot}),
as already mentioned in~(\ref{delV}), the symmetron potential energy is insufficient to drive cosmic acceleration.
Hence we will include a cosmological constant to reproduce the $\Lambda$CDM expansion history. Our main interest will be to
check that the symmetron ends up near the symmetry-breaking value by the present time, as assumed in the analysis of
spherically-symmetric solutions and tests of gravity. In Section~\ref{generalpot} we will extend the analysis to potentials
with greater promise of impacting both the recent expansion history and the growth history on linear scales today.

\subsection{Quartic Potential}
\label{quarticphasetran}

\begin{figure}[h]
\centering
\subfloat[]{\includegraphics[scale=1]{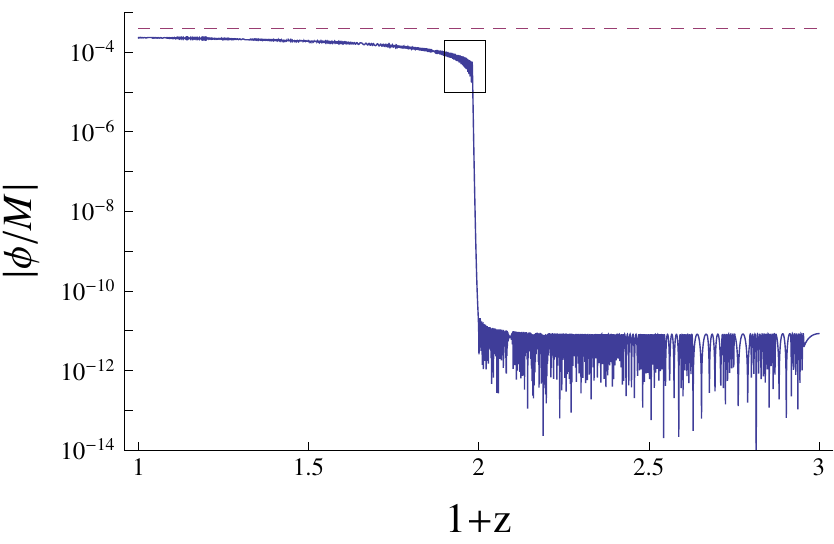}} \,
\subfloat[]{\includegraphics[scale=1]{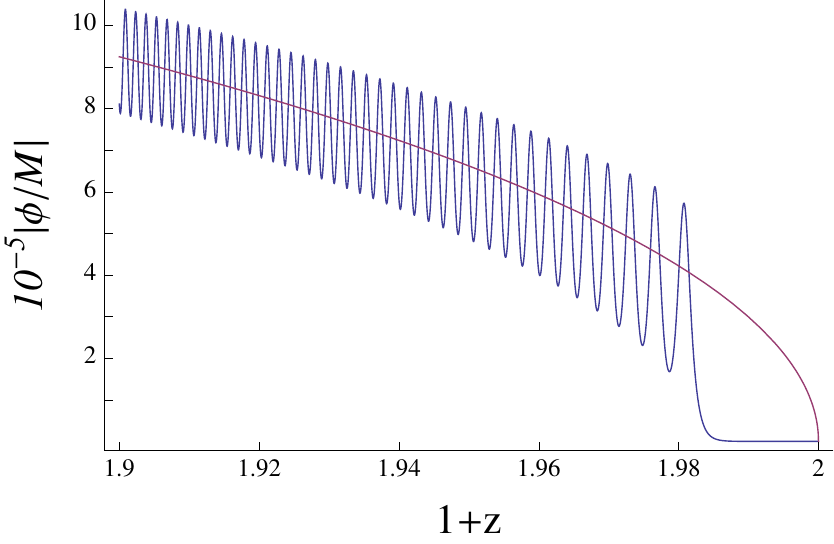}} 
\caption{Evolution of the symmetron around the phase transition. a) At the onset of the transition ($z_{\rm tran} = 1$), the symmetron rapidly evolves towards $\phi_0$,
the minimum of the potential, and undergoes damped oscillations around this point. The dashed line indicates $\phi_\mathrm{min}$, where we have chosen $M = 10^{-4}M_{\rm Pl}$.
b) Zoom-in on the boxed region, which shows that the oscillations are smooth. The solid line indicates $\phi_{\rm min}(z)$, the minimum of the effective potential given by~(\ref{phiminz}).}
\label{quartictran}
\end{figure}

For this analysis, we add a constant to the quartic potential~\ref{quarticpot},
\be
V(\phi) = -\frac{1}{2}\mu^2\phi^2 + \frac{1}{4}\lambda\phi^4 + \frac{\mu^4}{4\sqrt{\lambda}}\,, 
\label{quarticpotshifted}
\ee
such that the minima have zero potential energy. As shown in~(\ref{vev}), the minimum lies at $\phi_0 \sim M^2/M_{\rm Pl}\ll M$, hence we can still approximate the coupling function as quadratic: $A(\phi) \simeq 1 + \phi^2/2M^2$. We also include a cosmological constant, with energy density $\rho_\Lambda = 3H_0^2M_{\rm Pl}^2(1 - \Omega_{\rm m}^{(0)})$, where $\Omega_{\rm m}^{(0)}$ denotes as usual the present fractional energy density in matter. The scalar equation of motion~(\ref{scalarfrwfinal}) then takes the form
\be
\ddot{\phi} + 3H\dot{\phi} + \frac{3H_0^2M_{\rm Pl}^2}{M^2}\bigg(4 \left(1 - \Omega_{\rm m}^{(0)}\right) +  \Omega_{\rm m}^{(0)}(1+z)^3\bigg) \phi 
-\mu^2\phi + \lambda \phi^3 = 0\,.
\ee
The phase transition occurs at a redshift $z_{\rm tran}$ when the effective mass vanishes, that is,
\be
\mu^2 = \frac{3H_0^2M_{\rm Pl}^2}{M^2}\bigg(4\left(1-\Omega_{\rm m}^{(0)}\right) + \Omega_{\rm m}^{(0)}(1+z_{\rm tran})^3\bigg)\,.
\label{muonset}
\ee
For instance, the transition will happen at $z_{\rm tran} = 1$ for $\mu = \sqrt{15}H_0M_{\rm Pl}/M$, where we have substituted 
$\Omega_{\rm m}^{(0)} = 0.25$. This is consistent with~(\ref{muvalue}). 

Since $\mu \gg H_0$, shortly after the mass becomes tachyonic we expect the evolution of the symmetron to be adiabatic,
in the sense that the field should track the minimum of the effective potential:
\bea
\nonumber
\phi_{\rm min}(z) &=& \frac{1}{\sqrt{\lambda}}\left[\mu^2 -\frac{3H_0^2M_{\rm Pl}^2}{M^2}\bigg(4 \left(1 - \Omega_{\rm m}^{(0)}\right) +  \Omega_{\rm m}^{(0)}(1+z)^3\bigg)\right]^{1/2}\\
&=& \sqrt{3\Omega_{\rm m}^{(0)}} \frac{H_0M_{\rm Pl}}{\sqrt{\lambda}M} \bigg[ (1+z_{\rm tran})^3 - (1+z)^3\bigg]^{1/2}\,.
\label{phiminz}
\eea
This ``tracking" solution, valid for $z < z_{\rm tran}$, is an attractor because the effective potential is everywhere convex around this point.

To check these results, we have numerically integrated the combined equation~(\ref{scalarfrwfinal}) and~(\ref{friedeq}) from $z = 3$ until today,
including a matter component and a cosmological constant, with $\Omega_{\rm m}^{(0)} = 0.25$. For the quartic potential, we chose
$\mu = \sqrt{15}H_0M_{\rm Pl}/M$, corresponding to $z_{\rm tran} = 1$, and $\lambda = M_{\rm Pl}^4H_0^2/M^2$, consistent with~(\ref{lamvalue}).
As initial conditions for the symmetron, we choose $\phi(z =3) =  10^{-11}\,M$, which matches the example of Figure~\ref{numevol}.

\begin{figure}
\centering
\includegraphics[scale=1]{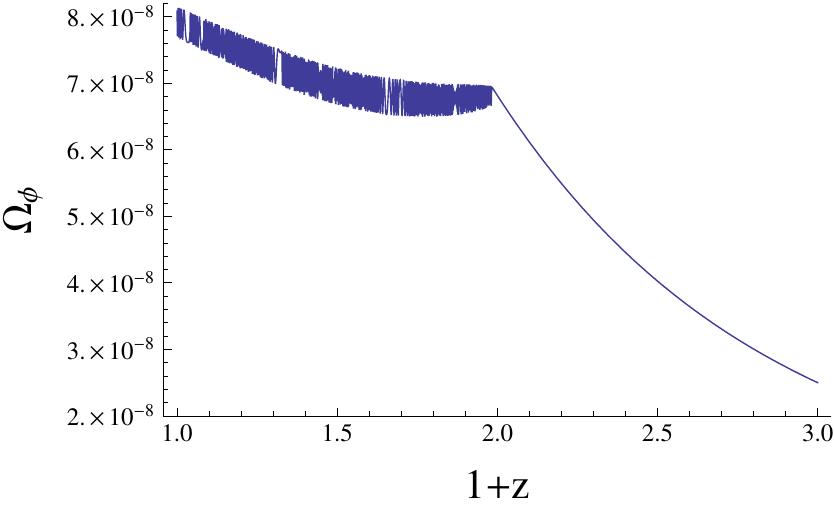}
\caption{The fractional energy density in the symmetron for the quartic potential.} 
\label{backreact}
\end{figure}

Figure~\ref{quartictran} shows the evolution of the symmetron field as a function of redshift.
At the onset of the transition, the symmetron rapidly rolls towards the minimum $\phi_0 = M^2/M_{\rm Pl}$ (denoted by the dashed line) and oscillates around this point. 
Figure~\ref{quartictran}b) zooms in on the oscillations in the boxed region, confirming that the symmetron nicely tracks $\phi_{\rm min}(z)$,
the minimum of the effective potential, as given by~(\ref{phiminz}).

Figure~\ref{backreact} shows the fractional energy density in the symmetron. This confirms that the symmetron backreaction is negligible,
as assumed in the analytical approximations above. The resulting expansion history is therefore indistinguishable from $\Lambda$CDM
in the quartic case.

\subsection{Generalized Potential and Coupling Function}
\label{generalpot}

\begin{figure}
\centering
\includegraphics[scale=1]{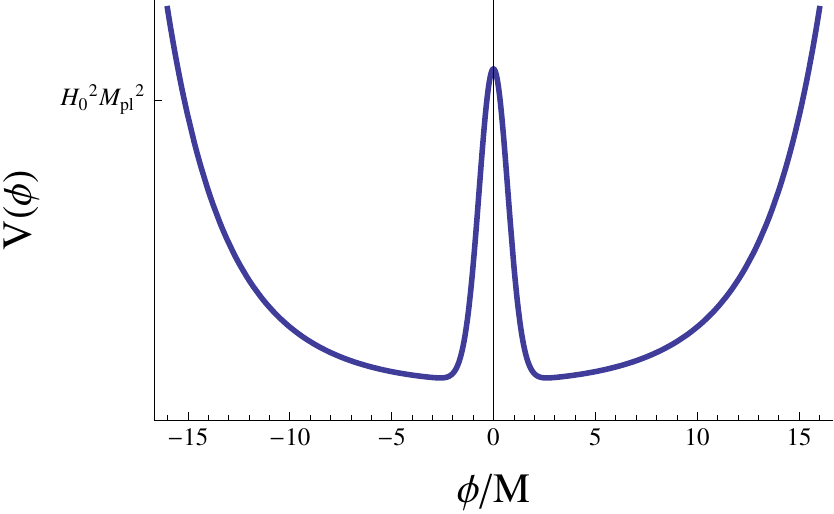}
\caption{The generalized potential studied in Section~\ref{generalpot}, shown here for illustrative purposes with $M =0.1M_{\rm Pl}$.
The minima occur at $\phi \sim \pm M$. By construction, the height of the potential is $\sim H_0^2M_{\rm Pl}^2$, and the mass around the minima is $\sim H_0$.}
\label{darkenergypot}
\end{figure}

To maximize the impact on the late-time expansion and growth histories, we generalize $V(\phi)$ such that the
height of the potential is comparable to the dark energy scale, $V_0 \sim H_0^2M_{\rm Pl}^2$, and the mass
around the minimum is of order Hubble, $m_0 \sim H_0$. The following general form satisfies these requirements: 
\be
V(\phi)= H_0^2 \Mpl^2 \left(e^{- \alpha\phi^2/2M^2}+\frac{M}{\Mpl}e^{\phi^2/2\Mpl^2}\right)\,,
\label{genpotential}
\ee
where $\alpha$ is a constant. This is sketched in Figure~\ref{darkenergypot}. For $|\phi|\ll M$,
\be
V(\phi) \simeq  H_0^2 \Mpl^2 - \frac{1}{2} \frac{\alpha H_0^2M_{\rm Pl}^2}{M^2}\phi^2 + \ldots \,,
\ee
this matches~(\ref{quarticpot}) to quadratic order, up to a constant, with $\mu = \sqrt{\alpha} H_0 M_{\rm Pl}/M$,
which is consistent with~(\ref{muvalue}). 

The minima in this case are located at values of order $M$,
\be
\phi_0 \simeq \sqrt{\frac{6}{\alpha}}M \ln^{1/2} \left(\frac{\alpha^{1/3}M_\mathrm{Pl}}{M}\right)\,.
\label{phi0gen}
\ee
Although one would have to carefully revisit the various tests of gravity analyzed in Section~\ref{tests},
a naive application of the thin shell condition $\Delta R_{\rm G}/R_{\rm G} = \phi_0/\Phi_{\rm G} M_{\rm Pl} \ll 1$ for our galaxy
would imply a tighter bound $M \,\lsim\, 10^{-6}M_{\rm Pl}$. We will keep this bound in mind, though our conclusions
will be insensitive to the precise choice of $M$.

Note that the potential evaluated at the minima is $V(\phi_0) \simeq H_0^2 M M_{\rm Pl} \ll H_0^2M_{\rm Pl}^2$,
hence the difference in potential energy with the maximum, $\Delta V \simeq H_0^2M_{\rm Pl}^2$, is of the order
of the dark energy scale, as desired. Meanwhile, the mass of small fluctuations around the minima is
\be
m_0 = H_0\sqrt{\frac{6 M}{M_{\rm Pl}}}\ln^{1/2} \left(\frac{\alpha^{1/3}M_\mathrm{Pl}}{M}\right)\,.
\ee
With $M = 10^{-6}M_{\rm Pl}$ and $\alpha = 1$, this gives $m_0 \simeq 10^{-2} H_0$, which is smaller than Hubble. Hence, as desired, the symmetron-mediated force extends to
the largest observable scales. Thus, as Figure~\ref{darkenergypot} illustrates, the potential
has a sharply-peaked maximum at $\phi = 0$ and shallow minima at $\phi = \pm \phi_0$.

Because $\phi$ reaches values of order $M$ in this case, the quadratic approximation~(\ref{Aphisimplify}) for $A(\phi)$
breaks down. Not only are corrections expected for $\phi \sim M$, they are in fact necessary to prevent the symmetron-mediated
force from being unacceptably strong. Indeed, if $A(\phi) = 1 + \phi^2/2M^2$ were valid full-stop, from~(\ref{forceratio}) the
force between two test masses in vacuum would be $F_\phi/F_{\rm N} \sim M_{\rm Pl}^2/M^2\sim 10^{12}$. Hence we must
generalize $A(\phi)$ by demanding that the scalar force be at most comparable to gravity for all
field values: ${\rm d}\ln A/{\rm d}\phi < 1/M_{\rm Pl}$ for all $\phi$. As a particular example, consider 
\be
A(\phi)=1+\frac{\phi^2}{2M^2 + M_{\rm Pl} \left|\phi\right|} \,.
\label{gencoupling}
\ee
This reduces to $A(\phi) \simeq 1 + \phi^2/2M^2$ for small $\phi$ and tends to $A(\phi) \simeq 1 + |\phi|/M_{\rm Pl}$
for large $\phi$. 

\begin{figure}
\centering
\includegraphics[scale=1]{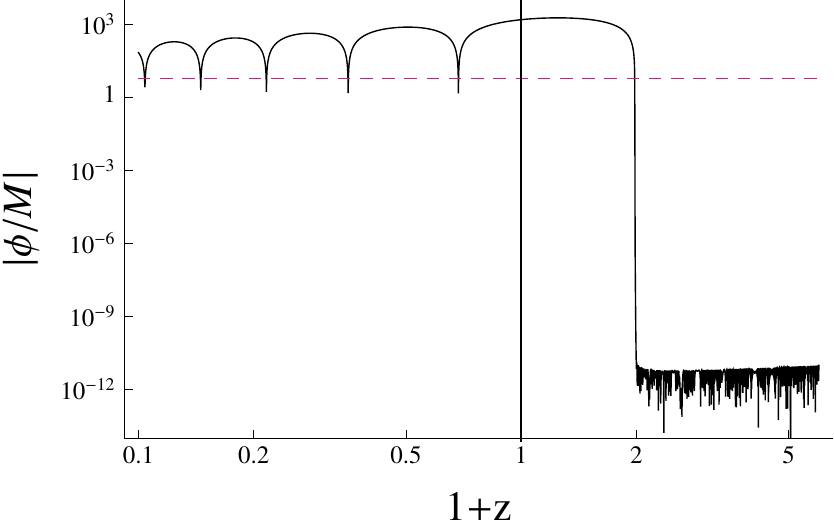}
\caption{The behavior of $\phi$ after the transition for the generalized potential~(\ref{genpotential}) and coupling function~(\ref{gencoupling}). The dashed line
denotes the minimum $\phi_0$, given by~(\ref{phi0gen}), where we have taken $M= 10^{-4}M_{\rm Pl}$. We have chosen parameters such that $z_{\rm tran} = 1$.
Note that the plot extends to future redshifts ($z< 0$). Because of the asymmetric form of the potential, in this case the symmetron has not yet settled to the minimum
by the present time.}
\label{symgen}
\end{figure}

With this choice of $V(\phi)$ and $A(\phi)$, we numerically solved the cosmological equations~(\ref{scalarfrwfinal})
and~(\ref{friedeq}). Since our goal is to have the symmetron drive cosmic acceleration, we
did not included a cosmological constant in this case. To facilitate the comparison with the quartic case of Section~\ref{quarticphasetran},
we chose $M = 10^{-4}M_{\rm Pl}$ and picked $\alpha$ so that the transition occurs at $z_{\rm tran} \simeq 1$.

The solution for the scalar field is shown in Figure~\ref{symgen}. Shortly after the onset of the phase transition,
we notice that the symmetron overshoots the minimum and reaches a value of order $M_{\rm Pl}$. Consequently, the field has not yet 
converged to $\phi_0$ by the present time --- note that the Figure even extends to $z < 0$ --- which is problematic
for tests of gravity. 

\begin{figure}
\centering
\includegraphics[scale=1]{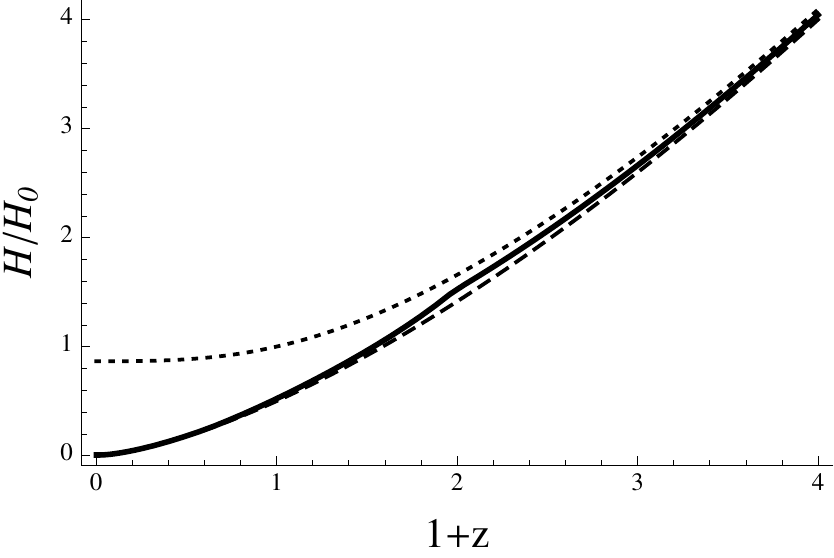}
\caption{Hubble parameter as a function of redshift for the symmetron with generalized potential~(\ref{genpotential}) and coupling function~(\ref{gencoupling}) (solid line),
where $H_0 = 72~{\rm km}\, {\rm s}^{-1}\,{\rm Mpc}^{-1}$ denotes the measured value. For comparison, the thick dashed line is a $\Lambda$CDM cosmology with
$\Omega_{\rm m}^{(0)} = 0.25$, while the thin dashed line is an Einstein de Sitter ($\Omega_{\rm m}^{(0)} = 1$) universe. In all three cases, we have fixed the
expansion history deep in the matter-dominated era.}
\label{cosmo}
\end{figure}

Figure~\ref{cosmo} shows the Hubble parameter $H(z)$ for the solution (solid line). For comparison,
the dotted line is a $\Lambda$CDM model with $\Omega_m^{(0)} = 0.25$, while the dashed line is an Einstein-de
Sitter ($\Omega_{\rm m}^{(0)} = 1$) universe. Aside from the small blip at $z \simeq z_{\rm tran}$, the solution
rapidly converges to the Einstein-de Sitter evolution, indicating that the symmetron energy density is negligible for
$z < z_{\rm tran}$. The failure to drive cosmic acceleration is due to the sharply peaked maximum, which leads to a rapid
completion of the transition and the field overshooting the minimum.

An obvious way to avoid the overshooting problem requires tuning parameters such that the transition
occurs very recently ($z_{\rm tran} \ll 1$). In this way, one can make the field value today of order $M$,
as the field evolves towards much larger values. At the same time, the potential energy at the maximum
can drive cosmic acceleration until the transition, matching the $\Lambda$CDM expansion history until
the very recent past. This solution, though consistent, is clearly not appealing as it requires
delicate tuning of parameters. To summarize, though we have focused on a particular class of potentials,
it appears difficult for the symmetron to drive cosmic acceleration while satisfying local tests of gravity, unless
one considers tuning parameters.

\section{The Inflationary Phase}
\label{inf}

Having studied the symmetron evolution throughout the radiation- and matter-dominated era, including the late-time symmetry-breaking
transition, we now go back and study more closely the issue of initial conditions deep in the radiation era. While the initial value
$\phi_{\rm rad-i}$ at the onset of the radiation-domination phase was treated as a free parameter in Section~\ref{cosmoevol},
in this Section we will see how this initial condition comes about from the field dynamics during inflation. We will find that the field
amplitude decays by an exponential amount during inflation, but can be exponentially resurrected during reheating. In particular, the condition
$\phi_{\rm rad-i} \ll M$ assumed in Section~\ref{cosmoevol} will translate to a constraint on the reheating phase. 

Of course, the inflationary analysis below must itself assume an initial value $\phi_{\rm inf-i}$, which depends on the pre-inflationary history.
To be conservative, however, we will set $\phi_{\rm inf-i} \simeq M$, corresponding to the largest field value consistent with the symmetron
energy density being subdominant during inflation.

\subsection{Dynamics During Inflation}

For simplicity, we treat the inflationary phase as exact de Sitter expansion, $a(t) \sim e^{H_{\rm inf}t}$, neglecting the backreaction
of the symmetron field. By virtue of its universal coupling to all matter fields, the symmetron couples to the inflaton, being sourced
by $T_{\rm inf} = (-1 + 3w_{\rm inf})\rho_{\rm inf} \simeq -4\rho_{\rm inf}$. 

The assumption that the symmetron is a spectator is of course not necessary --- though its backreaction could in principle
prevent inflation from starting --- but is made to simplify the analysis. In this case, as in the discussion of the radiation-dominated evolution,
for consistency the initial field value $\phi_{\rm inf-i}$ at the onset of inflation must satisfy $\phi_{\rm inf-i} \,\lsim\, M$, such that~(\ref{Aphisimplify})
applies. In particular, to a good approximation $A(\phi)\simeq 1$, and the inflaton contribution to the Friedmann equation~(\ref{friedeq})
is nearly $\phi$-independent and hence constant:
\be
3H_{\rm inf}^2M_{\rm Pl}^2\simeq \rho_{\rm inf}\,.
\ee
Meanwhile, the scalar equation of motion~(\ref{scalarfrwfinal}) reduces to
\be
\ddot{\phi}+3 H_{\rm inf} \dot{\phi}+12 \frac{M_\mathrm{Pl}^2 H^2_{\rm inf}}{M^2} \phi=0 \,.
\label{scalarinf}
\ee
Again, this is just the equation for a damped harmonic oscillator, with solution
\be
\phi =\phi_{\rm inf-i}e^{-3H_{\rm inf}t/2} \cos\left(\omega t \right) \,,
\label{Infsol}
\ee
where
\be
\omega=2\sqrt{3} \frac{\Mpl H_{\rm inf}}{M}\sqrt{1-\frac{9}{48}\(M\over \Mpl\)^2} \simeq 2\sqrt{3}  \frac{\Mpl H_{\rm inf}}{M}\,.
\label{omdef}
\ee
Clearly, $\omega \gg H_{\rm inf}$, and the symmetron oscillates rapidly as inflation proceeds. The choice of phase in~(\ref{Infsol})
corresponds to the symmetron starting nearly from rest from $\phi_{\rm inf-i}$ at the onset of inflation ($t=0$).

Since the effective symmetron mass is large compared to Hubble, its energy density averaged over many oscillations redshifts as
dust:
\be
\langle \rho_\phi \rangle \simeq \frac{1}{2} \omega^2 \phi_{\rm inf-i}^2 e^{-3H_{\rm inf}t}  = 6 H_{\rm inf}^2 M_{\rm Pl}^2 \left(\frac{\phi_{\rm inf-i}}{M}\right)^2 e^{-3H_{\rm inf}t} \,.
\label{rhoredshift}
\ee
This confirms that $\phi_{\rm inf-i} \ll M$ is indeed necessary to neglect the symmetron backreaction from the outset,
an approximation which only gets better in time, as~(\ref{rhoredshift}) implies.
Over the entire period of inflation, therefore, the field amplitude decays by
\be
\frac{|\phi_{\rm inf-end}|}{|\phi_{\rm inf-i}|} =  e^{-3N_{\rm inf}/2}\,,
\label{infdamping}
\ee
where $N_{\rm inf}$ is the number of inflationary e-folds. For the standard value $N_{\rm inf} = 60$, corresponding
to GUT-scale inflation, this gives a suppression factor of $e^{-90} \simeq 10^{-39}$. 

\subsection{Reheating}

We next track the evolution of the symmetron during reheating, as the inflaton decays into radiation.
Once again neglecting the symmetron backreaction, the scalar equation~(\ref{scalarfrwfinal}) becomes
\be
\ddot{\phi} + 3H\dot{\phi}   + \frac{\rho}{M^2} (1 - 3w) \phi  = 0\,,
\label{scalarreheat0}
\ee
where we have also neglected the potential term $V_{,\phi}$. Here $\rho$ and $w$ respectively denote the energy
density and equation of state of the combined inflaton-radiation fluid. To simplify the analysis, we will assume that reheating
occurs rapidly compared to a Hubble time, so that $H \simeq H_{\rm inf}$ remains approximately constant. Moreover, in this regime
the Hubble friction term can be ignored, and~(\ref{scalarreheat0}) reduces to
\be
\ddot{\phi} +  \frac{\omega^2}{4}(1 - 3w) \phi = 0\,,
\label{scalarreheat1}
\ee
where we have used $\rho  = 3H^2M_{\rm Pl}^2 \simeq 3H_{\rm inf}^2M_{\rm Pl}^2$, and where $\omega \simeq 2\sqrt{3}M_{\rm Pl}H_{\rm inf}/M$
was defined in~(\ref{omdef}).

To proceed, we need to specify $w(t)$ during reheating. For concreteness, we consider the simplest case of the inflaton decaying perturbatively
into radiation as it oscillates around the minimum of its potential. In the process, $w$ oscillates between $-1$ and $+1$, and approaches $1/3$ 
as the universe becomes dominated by radiation. To model this evolution, consider 
\be
w(t) = \frac{1}{3} \left[ 1 - e^{-\Gamma_{\rm inf} t}\left( 1 + 3\cos m_{\rm inf}t \right)\right]\,,
\label{wN}
\ee
where $t = 0$ now marks the onset of the reheating phase. This form is such that the equation of state starts out with $w(0) = -1$, thereby matching smoothly to the inflationary phase. At late times,
$w \rightarrow 1/3$, as desired. Manifestly, $\Gamma_{\rm inf}$ and $m_{\rm inf}$ denote the inflaton decay rate and mass, respectively. 

%%% FIGURE
\begin{figure}
\centering
\subfloat[$\Gamma_{\rm inf} = 10\,\minf$ and $m_{\rm inf} = \omega$.]{\includegraphics[scale=1.0]{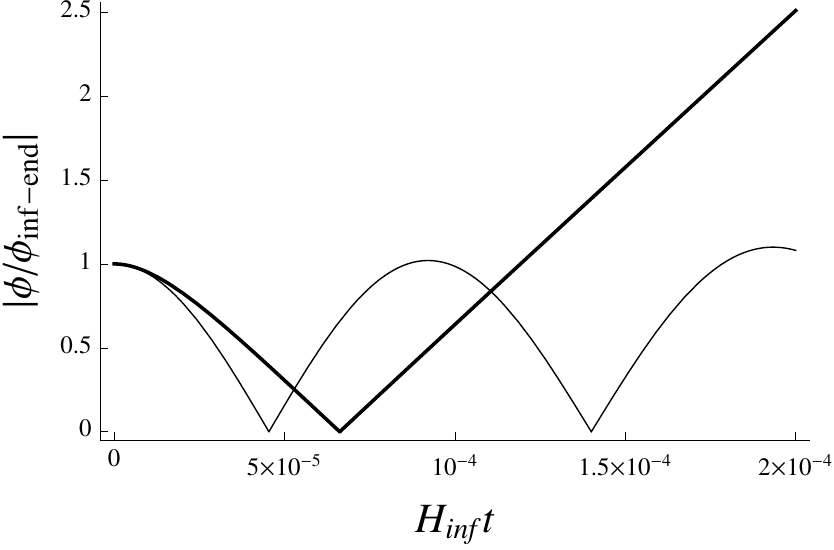}}
\;
\subfloat[$\Gamma_{\rm inf} =3.2\times 10^{-2}m_{\rm inf}$ and $m_{\rm inf} = \omega$.]{\includegraphics[scale=1.0]{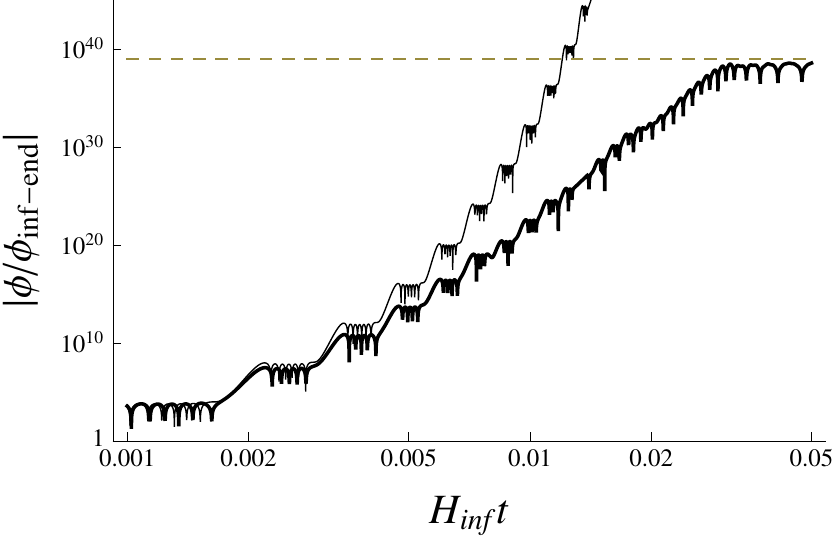}} \\
\subfloat[$\Gamma_{\rm inf} = 10^{-4}m_{\rm inf}$ and $m_{\rm inf} = 10^{3} \omega$.]{\includegraphics[scale=1.0]{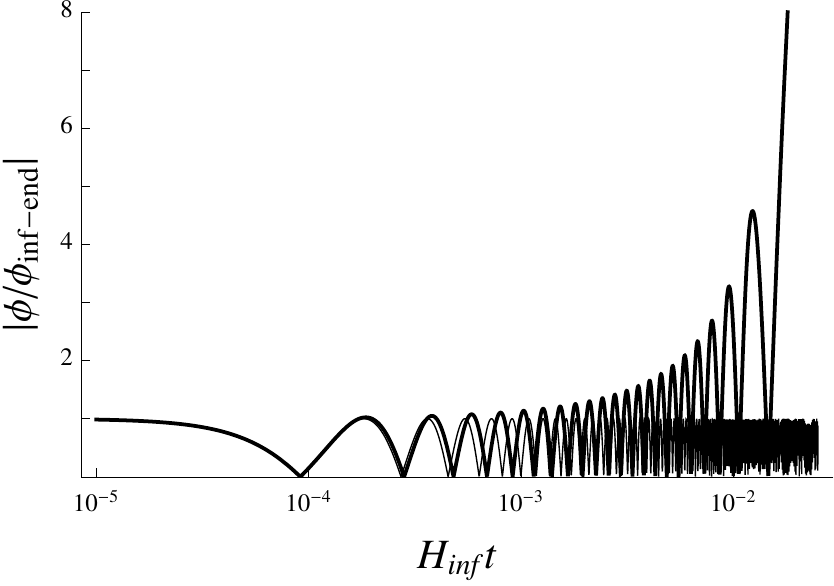}} 
\caption{Symmetron evolution during reheating for different parameter values. In each case, the thick line indicates the numerical solution,
whereas the thin line is the Mathieu cosine function~(\ref{mathieucosine}). All plots have $\omega = 10^4H_{\rm inf}$ and $M = 10^{-4}M_{\rm Pl}$.
a) In this case, $\Gamma_{\rm inf} \gg m_{\rm inf}$, hence reheating has essentially completed before the tachyonic amplification can kick in. b) In the opposite regime, $\Gamma_{\rm inf} \ll m_{\rm inf}$, reheating is slow. Since $m_{\rm inf} = \omega$ lies in the unstable region of the Mathieu cosine solution, the field is exponentially amplified. The amplification is
eventually curtailed once the inflaton decays and reheating takes place. The dashed line indicates the maximum allowed amplification, corresponding to $\phi_{\rm rad-i} \simeq M$.
c)  This case also has  $\Gamma_{\rm inf} \ll m_{\rm inf}$, as in b), but $m_{\rm inf} = 10^{3} \omega$ lies in the stability region of the Mathieu cosine. Note the logarithmic scale on the time axis in this case.}
\label{reheatamp}
\end{figure}
%%% END FIGURE

With this choice of $w$,~(\ref{scalarreheat1}) becomes
\be
\ddot{\phi} +   \frac{\omega^2}{4} e^{-\Gamma_{\rm inf} t} \left(1 + 3\cos m_{\rm inf} t \right) \phi = 0\,.
\label{scalarreheat}
\ee
Thus the evolution is characterized by three dimensionful parameters: the inflaton decay rate $\Gamma_{\rm inf}$, the inflaton mass $m_{\rm inf}$,
and the effective symmetron mass $\omega$ during inflation. Since reheating is assumed to proceed rapidly on a Hubble time, all three parameters
are $\gg H_{\rm inf}$. Whenever the equation of state achieves $w > 1/3$, the effective symmetron mass term in~(\ref{scalarreheat}) becomes
tachyonic, which can amplify the field value. As $w \rightarrow 1/3$, the effective mass is driven to zero, and the field 
evolves freely. Eventually the field will be slowed down by Hubble friction. 

The behavior of the solution can be understood by first considering the regime $t\ll \Gamma_{\rm inf}^{-1}$,
where the exponential damping term is nearly constant. Defining $x \equiv m_{\rm inf}t/2$ and $y \equiv \phi/\phi_{\rm inf-end}$,
this scalar equation reduces to a Mathieu equation
\be
\frac{{\rm d}^2 y}{{\rm d}x^2} + \left[ a - 2q\cos 2x \right]y = 0\,,
\ee
with
\be
a \equiv \frac{\omega^2}{m_{\rm inf}^2}\;;\qquad q = -\frac{3}{2}a\,.
\ee
The unique solution with $y(0) = 1$ and ${\rm d}y/{\rm d}x(0) = 0$ is the Mathieu cosine function, $y = C(a,q,x)$,
which in our case is given by
\be
\phi (t) \simeq \phi_{\rm inf-end}\, C\left(a,-\frac{3}{2}a, \frac{m_{\rm inf} t}{2}\right) \qquad \;\;\;\;\;\; {\rm for}\;\; t\ll \Gamma_{\rm inf}^{-1}\,.
\label{mathieucosine}
\ee
The stability of the symmetron evolution is therefore determined by the stability properties of the Mathieu cosine function.
For $a \,\lsim\, 10^{-2}$, corresponding to $m_{\rm inf}\,\gsim\, 10\,\omega$, we find that the solution undergoes stable oscillations, consistent with the fact that
$C(a,q,x) \approx \cos(\sqrt{a} x)$ for $|q|\ll 1$. For $a \,\gsim\, 10^{-2}$, corresponding to $m_{\rm inf}\,\lsim\, 10\,\omega$, the solution is unstable, and the
symmetron amplitude grows exponentially. The growth is eventually cut off when $t \sim \Gamma_{\rm inf}^{-1}$, and the source becomes exponentially suppressed,
indicating that reheating has nearly completed.

Figure~\ref{reheatamp} confirms these expectations. In all cases, the thick line is the numerical solution, 
while the thin line is the Mathieu cosine function~(\ref{mathieucosine}). Figure~\ref{reheatamp}a) is an example
of rapid reheating,  $\Gamma_{\rm inf} \gg m_{\rm inf}$. In this case, reheating is essentially complete before the
tachyonic amplification can kick in, and the Mathieu cosine is a good approximation for less than one symmetron
oscillation. 

Figures~\ref{reheatamp}b) and c) are examples of slow reheating, $\Gamma_{\rm inf} \ll m_{\rm inf}$, in which case
the Mathieu cosine matches the exact solution for many oscillations. Figure~\ref{reheatamp}b) corresponds to $m_{\rm inf} = \omega$,
which lies within the unstable region of the Mathieu cosine parameter space. The field is amplified by many orders of magnitude, until 
its growth is tamed after a time $t \sim \Gamma_{\rm inf}^{-1}$ when reheating takes place. The dashed line indicates the maximum allowed amplification
so that $|\phi_{\rm rad-i}| < M$, as discussed shortly. The parameters have been chosen in this case such that the amplification barely satisfies this bound.
Figure~\ref{reheatamp}c) corresponds to $m_{\rm inf} =  10^{3}\omega$, which lies within the stable region of the Mathieu cosine. Note that the scale on the $t$-axis is logarithmic in this case.

After several decay times, $\Gamma_{\rm inf} t\gg 1$, the driving term becomes negligible, and~(\ref{scalarreheat})
reduces to $\ddot{\phi}\simeq 0$. Thus the field eventually grows linearly with time, as confirmed
in all examples of Figure~\ref{reheatamp}. Of course this linear growth eventually gets cut off by Hubble friction. 
The asymptotic value is  is identified with $\phi_{\rm rad-i}$, the initial value for the radiation-dominated evolution.

We can derive a constraint on the reheating parameters by demanding that $\phi_{\rm rad-i} < M$,
for consistency of neglecting the symmetron backreaction, as assumed throughout Section~\ref{RDera}. For simplicity, we estimate 
$\phi_{\rm rad-i}$ as the field value at $t\sim \Gamma_{\rm inf}^{-1}$, and ignore the relatively small subsequent growth.
To be most conservative, we take $\phi_{\rm inf-i} = M$ at the onset of inflation. Assuming $N_{\rm inf} = 60$ e-folds of
inflation,~(\ref{infdamping}) implies $\phi_{\rm inf-end} \simeq 10^{-39} M$. The condition $\phi_{\rm rad-i} < M$ therefore corresponds to
\be
\left\vert\frac{\phi \left(t = \Gamma_{\rm inf}^{-1}\right)}{\phi_{\rm inf-end}}\right\vert \,\lsim\, 10^{39}\,.
\label{infbound}
\ee
As discussed above, there is minimal growth if $\Gamma_{\rm inf} \gg m_{\rm inf}$ (in which case reheating happens quickly)
and/or $m_{\rm inf}\,\gsim\, 10\,\omega$ (in which case the Mathieu cosine is stable). In the regime $\Gamma_{\rm inf} \ll m_{\rm inf}$
and $m_{\rm inf}\,\lsim\, 10\,\omega$, however, we have found numerically that~(\ref{infbound}) is satisfied for 
\be
\Gamma_{\rm inf} \, \gsim \,  0.025\,m_{\rm inf} \left(\frac{\omega}{m_{\rm inf}}\right)^{0.7}\,.
\label{Gammabound}
\ee
To summarize, due to possible parametric resonance during reheating, the symmetron amplitude can increase exponentially.
Demanding that this amplification remains within bounds imposes constraints on the parameters, which we have derived in the limit
that all time scales involved are short relative to a Hubble time.

 \section{\label{conclsec}Conclusions}

Despite the overwhelming evidence for the existence of dark energy and dark matter, little is known
about their underlying fundamental physics. Over the last few years there has been
considerable activity exploring the possibility that the dark sector includes new light degrees of freedom
(generally scalar fields) that couple not only to dark matter but also to ordinary (baryonic) matter. 
Much of the research efforts have focused on the development of {\it screening mechanisms} to explain why such scalars, if light,
have escaped detection from laboratory/solar system tests of gravity. The manifestation of these scalar fields therefore
depends sensitively on their environment, which in turn leads to striking experimental signatures.

One such screening mechanism is the symmetron mechanism, which relies on a scalar field having a VEV that depends on the ambient matter density.
The VEV is small in regions of high density, and large in regions of low density. In addition, the coupling of the scalar to matter
is proportional to this VEV, hence the scalar couples with gravitational strength in regions of low density, but couples much more weakly
in regions of high density. 

In this paper we have derived the cosmological expansion history in the presence of a symmetron field,
tracking the evolution of the scalar through the inflationary epoch, the phase of reheating, the standard
radiation- and matter-dominated eras, and through the late-time symmetry-breaking phase transition.
For a wide range of initial field values at the onset of inflation, we have shown that the symmetron
ends up in the symmetry-breaking vacuum by the present time, as assumed in the derivation of static,
spherically-symmetric solutions and analysis of tests of gravity.

For the fiducial quartic potential, the scales involved are too small to drive late-time cosmic acceleration,
hence we included a cosmological constant. The symmetron backreaction is consistently small throughout
the evolution, hence the expansion history is consistent with observations. We also introduced a general
class of potentials, engineered so that the potential energy difference is of order the present critical density,
and the mass around the minimum is of order Hubble today. In the simplest examples considered here,
however, the symmetron was unsuccessful in driving cosmic acceleration.

Our results form the foundation of future investigations of the cosmological implications of symmetrons.
The natural next step is to study the evolution of density perturbations in the presence
of a symmetron, both in terms of the linear growth history and in the non-linear regime using
N-body simulations. As in the chameleon and Vainshtein screening mechanisms, the additional
scalar force should enhance structure growth at late times. It would be interesting to contrast the impact on 
structure formation among different screening mechanisms and seek distinguishing signatures.

Another cosmological probe of symmetron physics is the production of topological domain walls. 
In this paper we have circumvented this fascinating issue by implicitly focusing on initial conditions
where the scalar field ends up in the vicinity of $\phi = 0$ at late times, but sufficiently displaced such
that the phase transition can proceed homogeneously. More generally, we expect the symmetron
to result in late-forming domain wall networks. It will be interesting to study their formation and 
contrast the phenomenology of symmetron defects with standard topological defect networks
by comparing their observational impact, {\it e.g.}, on CMB anisotropy, structure growth and gravitational lensing.

\textit{Acknowledgments:} It is our pleasure to thank Joseph Clampitt, Lam Hui, Bhuvnesh Jain and Junpu Wang for many helpful discussions.
This work was supported in part by College Alumni Society Board of Managers and Presidents Undergraduate Research Grant (A.L.), 
a University Research Foundation grant from the University of Pennsylvania, NSF grant PHY-0930521, and the Alfred P. Sloan Foundation (J.K.).

\bibliographystyle{utphys}
%\addcontentsline{toc}{section}{References}
\bibliography{symmetroncosmology1}

\end{document}